\documentclass[prb,preprint,amsmath,amssymb,superscriptaddress]{revtex4} 

\usepackage{graphicx}
\usepackage{bm}
\usepackage{amsmath}
\usepackage{amsfonts}
\usepackage{amssymb}

\begin{document}

\title{Transverse optical instability patterns in semiconductor
microcavities: polariton scattering and low-intensity all-optical switching}

\author{M.H. Luk}

\affiliation{Department of Physics and Institute of Theoretical
Physics, The Chinese University of Hong Kong, Hong Kong SAR, China}

\author{Y.C. Tse}

\affiliation{Department of Physics and Institute of Theoretical
Physics, The Chinese University of Hong Kong, Hong Kong SAR, China}

\author{N.H. Kwong}

\email{kwong@optics.arizona.edu}

\affiliation{Department of Physics and Institute of Theoretical
Physics, The Chinese University of Hong Kong, Hong Kong SAR, China}

\affiliation{Center of Optical Sciences, The Chinese University of
Hong Kong, Hong Kong SAR, China}

\affiliation{College of Optical Sciences, University of Arizona,
Tucson, AZ 85721, USA}

\author{P.T. Leung}

\affiliation{Department of Physics and Institute of Theoretical
Physics, The Chinese University of Hong Kong, Hong Kong SAR, China}

\author{Przemyslaw Lewandowski}

\affiliation{Physics Department and Center for Optoelectronics and
Photonics Paderborn (CeOPP), Universit\"at Paderborn, Warburger
Strasse 100, 33098 Paderborn, Germany}



\author{R. Binder}

\affiliation{College of Optical Sciences, University of Arizona,
Tucson, AZ 85721, USA}

\affiliation{Department of Physics, University of Arizona, Tucson,
AZ 85721, USA}

\author{Stefan Schumacher}

\affiliation{Physics Department and Center for Optoelectronics and
Photonics Paderborn (CeOPP), Universit\"at Paderborn, Warburger
Strasse 100, 33098 Paderborn, Germany}

\date{\today}

\begin{abstract}
We present a detailed theoretical study of transverse
exciton-polariton patterns in semiconductor quantum-well
microcavities. These patterns are initiated by directional
instabilities (driven mainly by polariton-polariton scattering) in
the uniform pump-generated polariton field and are measured as
optical patterns in a transverse plane in the far field. Based on a
microscopic many-particle theory, we investigate the spatio-temporal
dynamics of the formation, selection, and optical control of these
patterns. An emphasis is placed on a previously proposed
low-intensity, all-optical switching scheme designed to exploit
these instability-driven patterns. Simulations and detailed analyses
of simplified and more physically transparent models are used. Two
aspects of the problem are studied in detail. First, we study the
dependencies of the stability behaviors of various patterns, as well
as transition time scales, on parameters relevant to the switching
action. These parameters are the degree of built-in azimuthal
anisotropy in the system and the switching (control) beam intensity.
It is found that if the parameters are varied incrementally, the
pattern system undergoes abrupt transitions at threshold parameter
values which are accompanied by multiple-stability and hysteresis
behaviors. Moreover, during a real-time switching action, the
transient dynamics of the system, in particular the transition time
scale, may depend significantly on the proximity of unstable
patterns. The second aspect is a classification and detailed
analysis of the polariton scattering processes contributing to the
pattern dynamics, giving us an understanding of the
selection and control of patterns as results of these processes'
intricate interplay. The crucial role played by the (relative)
phases of the polariton amplitudes in determining the gains and/or
losses of polariton densities in various momentum modes is
highlighted. As a result of this analysis, an interpretation of the
actions of the various processes in terms of concepts commonly used
in classical pattern-forming systems is given.

\end{abstract}

\maketitle

\section{Introduction}
\label{intro.sec}

Laser beams propagating through a nonlinear medium can under certain
conditions undergo directional instabilities. These instabilities
are driven by phase-conjugate wave-mixing
processes,
\cite{Yariv1977,Grynberg1989,dalessandro-firth.91,Geddes1994,lega-etal.95,
lugiato-etal.99,Kheradmand2008,schumacher-etal.09}
leading to spontaneous generation and build-up of intensity in modes
with propagation directions other than those of the
input beams. The system's transverse translational symmetry is thus
spontaneously broken, resulting in periodic or quasiperiodic
modulational patterns, e.g. stripes and hexagons, in the intensity profile. These spontaneously
formed optical patterns have been observed in experiments with
lasers, \cite{hegarty-etal.99,houlthan-etal.01} and such passive
optical systems as atomic gases,\cite{grynberg-etal.88,giusfredi-etal.88,herrero-etal.99,Dawes2005},
liquid crystal light valves, \cite{benkler-etal.00}, photorefractive crystals,\cite{denz-etal.98} and, most recently, semiconductor quantum-well
microcavities.\cite{ardizzone-etal.13} (For general reviews, see e.g.
Refs.~\onlinecite{abraham-firth.90,lugiato.94,staliunas-sanchez.03}.)
In a broader context, these developments are instances of
spontaneous pattern formation found in a wide range of
nonequilibrium physical systems such as classical fluids and chemical reaction systems.
\cite{cross-hohenberg.93,newell-etal.93}

The occurrence of directional instability patterns in optics has
also indicated the potential of using them for low-intensity
all-optical switching. It was demonstrated in
Ref.~\onlinecite{Dawes2005} that the orientation of transverse
patterns in a rubidium vapor can be all-optically and reversibly
switched by the application of a weak control beam. Subsequently, it
was proposed that a similar switching scheme can also be implemented
in a semiconductor quantum-well
microcavity.\cite{schumacher-etal.09} Another analysis also shows
the possibility to optically control transverse patterns in planar
semiconductor structures.\cite{Kheradmand2008} A review of these
activities was given in Ref.~\onlinecite{Dawes2009}. The switching scheme devised in Ref.~\onlinecite{schumacher-etal.09} has recently been experimentally demonstrated in a quantum-well double-cavity system.\cite{ardizzone-etal.13} In this
context, it is worthwhile to note the different physical origins of pattern
formation in these systems. In an optically excited
atomic vapor,\cite{Dawes2005} nonlinearities stem from partial
saturation of the atomic resonances, and three-dimensional
phase-matching conditions have to be satisfied with
counterpropagating pump beams for efficient off-axis four-wave
mixing of light to take place. In contrast, in a planar quantum-well
based microcavity, nonlinearities arise from the parametric scattering of
exciton-polaritons\cite{Savvidis2000,Huang2000,Savasta2003} (polariton
four-wave mixing), which is predominantly driven by the Coulomb interaction
between the polaritons' excitonic
constituents,\cite{kwong-etal.01prb,takayama-etal.02,ciuti2000,Whittaker2001} with additional smaller
contributions from phase-space filling. Two-dimensional
phase-matching, in the cavity's plane, has to be fulfilled to render
the scattering processes efficient.

\begin{figure}[b]
\includegraphics
[scale=0.25, angle=0]{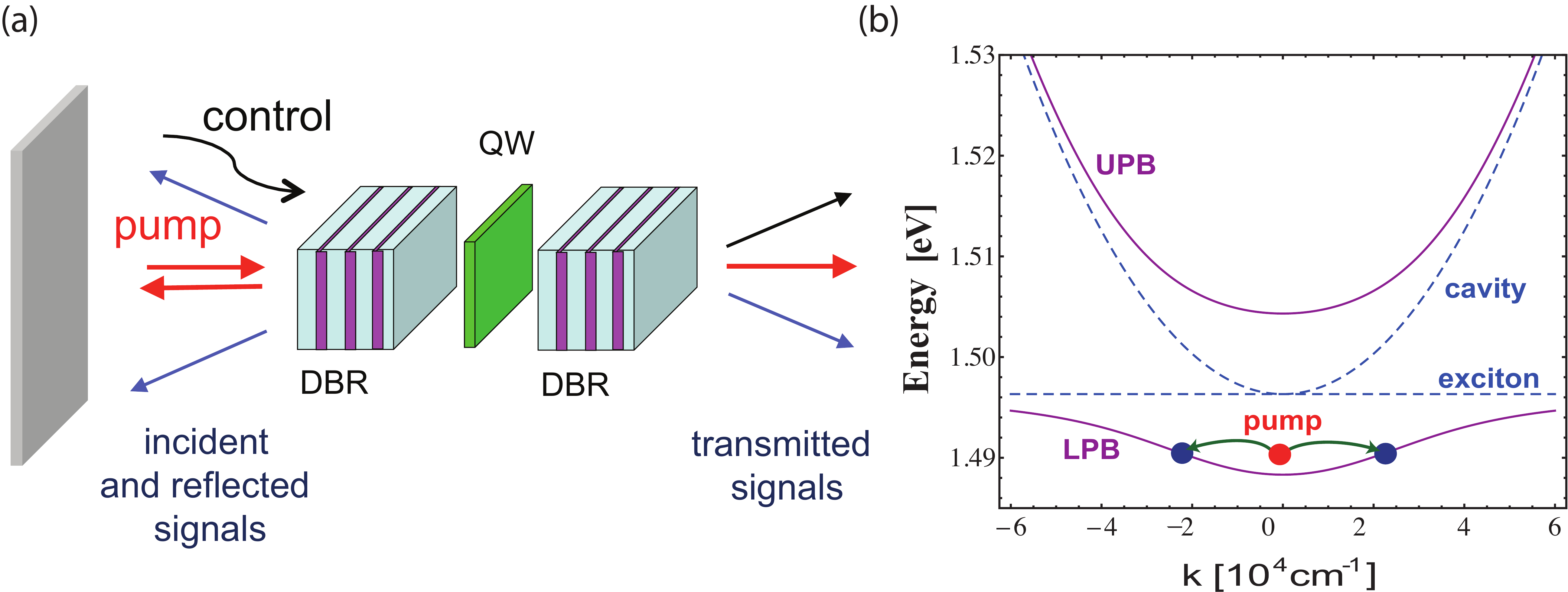}
\caption{(Color online.) (a) Sketch of the semiconductor
quantum-well microcavity. Two distributed Bragg reflectors (DBR)
form a photonic cavity with an embedded semiconductor quantum well
(QW). The pump is in normal incidence to the quantum-well plane.
When transverse instabilities are present, off-axis patterns can be
observed in the far field. The orientations of these patterns can be
rotated by applying a weak control beam. (b) Cavity polariton energy
vs in-plane momentum. The uncoupled exciton and cavity photon modes
are shown as dotted and dashed lines, respectively. The solid lines
represent the upper (UPB) and the lower (LPB) polariton branches. The displayed dispersion relations do not include the effects of (nonlinear) exciton interactions.
The pump frequency and scattering of two pump-induced polaritons to
off-axis modes are also indicated.}\label{sketch.fig}
\end{figure}

\begin{figure}[b]
\includegraphics[scale=0.5,angle=0,trim=00 00 00
00]{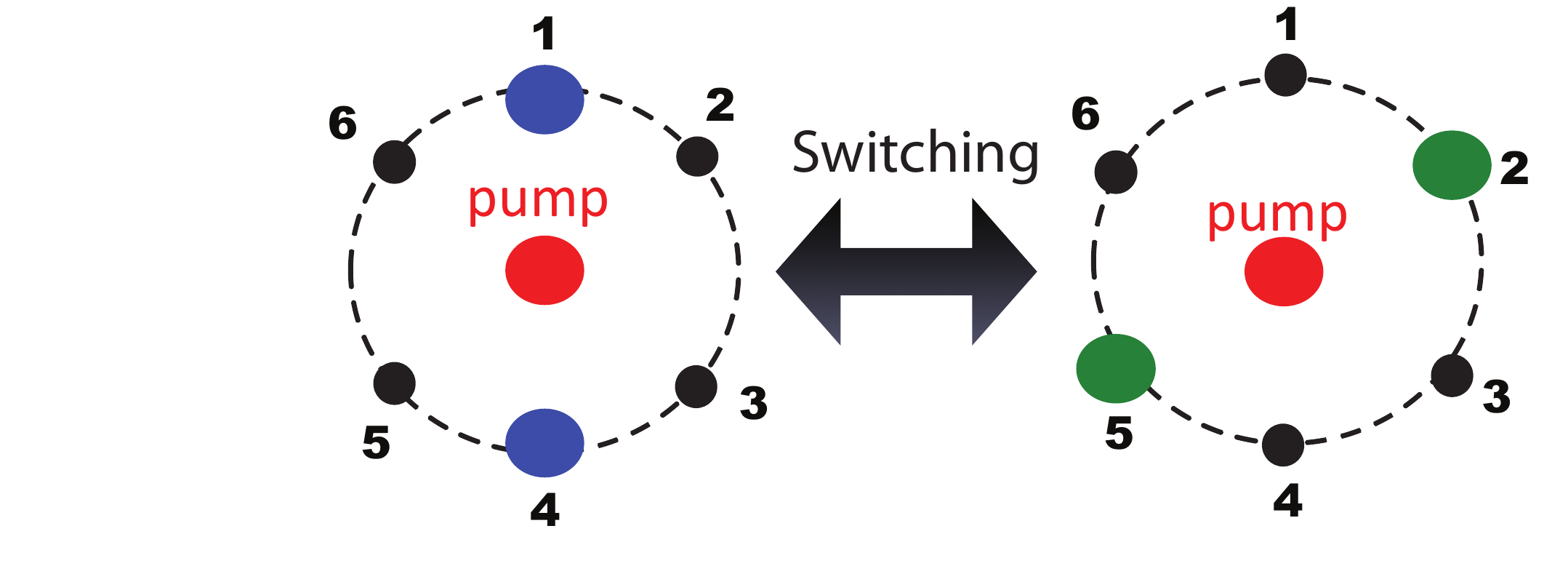} \caption{(Color online.) Schematic representation
of the switching process. The sites labeled 1 to 6 represent the
projected positions on a transverse plane in the far field of six
off-axis momentum modes, forming a hexagon.
On the left is shown an optical pattern with the mode pair 1 and 4
being spontaneously generated by phase-conjugate instabilities and
favored by a built-in anisotropy of the system. When a control beam
is directed at mode 2, the pattern is rotated to directions 2 and 5.
When the control is then turned off, the pattern reverts to its
original orientation.}\label{switch.fig}
\end{figure}

While setting the conceptual foundation of the
directional switching action in semiconductor microcavities in
Ref.~\onlinecite{schumacher-etal.09} and implementing the switching in Ref.~\onlinecite{ardizzone-etal.13} were important steps forward,
further work is required to fully understand the
underlying complicated dynamics. In this paper, we study this
problem theoretically as an instance of competitions among various
transverse patterns formed by the nonlinear interactions of exciton
polaritons. The switching action of the control beam is viewed as
destabilizing the initial pattern while stabilizing another. Two
aspects of this investigation are reported in this paper. First, we
study the dependencies of the stability behaviors of various
patterns,
as well as transition time scales, on
parameters relevant to the switching action. These parameters are
the degree of built-in azimuthal anisotropy in the system and the
control beam intensity. Second, we analyze in detail the actions of
physical processes contributing to the dynamics and form an
understanding of the selection and control of patterns as results of
these processes' intricate interplay.

The nonlinear optical and many-boson aspects of polaritons in semiconductor microcavities are an active area of current research.\cite{sanvitto-timofeev.12} Besides modulational intensity patterns, more complicated spatial instability structures such as solitons \cite{amo-etal.11} and vortices \cite{tosi-etal.12} are also being investigated in these systems. Optical parametric oscillation in various pump/signal configurations have been demonstrated in quantum well microcavities.\cite{stevenson-etal.00,Romanelli2007,ballarini-etal.09} This process shares a common underlying mechanism -- parametric polariton scattering -- with the directional instabilities studied here.

The quantum well microcavity configuration used in this paper is sketched in
Fig.~\ref{sketch.fig}(a). A photonic cavity is formed by a pair of
distributed Bragg reflectors (DBRs), and a semiconductor quantum
well (QW) is placed inside the cavity. One cavity photon mode is tuned to
coincide with the lowest 1s heavy hole excitonic resonance, leading
to strong coupling between the two and the formation of an upper
(UPB) and a lower (LPB) polariton branch. The dispersion relations
(energy vs in-plane momentum) of these linear optical excitations
for the parameters used in this paper are plotted in
Fig.~\ref{sketch.fig}(b). A pump beam comes in at normal incidence
to the cavity's plane and is spectrally tuned above the bottom of
the lower polariton branch but well below the bare exciton
resonance. It excites (virtual) polaritons with zero in-plane
momenta, pairs of which can scatter near-resonantly into modes with
finite, opposite transverse (in-plane) momenta on the LPB, as
indicated in Fig.~\ref{sketch.fig}(b).  The scattered polaritons
have a certain probability to emerge from the microcavity as
photons. Under favorable
conditions, when the pump intensity is above a threshold, the
emerging photons in the off-axis directions become intense and
coherent beams, characteristic of parametric, directional
instabilities of the pump beam. The instability reduces the system's symmetry from transverse translational to azimuthal. This azimuthal symmetry may be further broken by the nonlinear interactions among the off-axis polariton field amplitudes, thereby favoring patterns with
density in a finite number of clusters of $\bf k$ modes (i.e. modes in transverse momentum space), e.g. a hexagon or its subsets. Schematically illustrated in
Fig.~\ref{switch.fig}, the low-intensity switching scheme proposed in
Ref.~\onlinecite{schumacher-etal.09} involves transitions between
two two-spot patterns residing on the same hexagon. (Note the role of the built-in azimuthal
anisotropy in the preparation of the initial pattern.) Instead of
coherent polariton scattering, the instabilities may equivalently be
interpreted as arising from phase-conjugate four wave mixing of the
coherent polariton field.

The conditions
for the formation of hexagons and two-spot patterns are quite well
understood in other disciplines of physics.\cite{cross-hohenberg.93}
However, the additional degrees of complexity, anisotropy and
external real-time control, make pattern formation in our system,
and in the system in Ref.~\onlinecite{Dawes2005}, a much more
involved issue than found in commonly studied systems. Moreover,
unlike macroscopic pattern-forming systems, where the field is
real-valued, our polariton field is quantum mechanical and therefore
complex-valued, and the mutual activating and inhibiting actions of
the mode densities on each other are entwined with their relative
phases. This makes building physical intuition about these actions
and hence how patterns are selected a non-trivial exercise.

Our theoretical approach is based on a microscopic theory of
excitonic optical response of the semiconductor quantum
well \cite{takayama-etal.02,kwong-etal.01prb} coupled to a classical
treatment of the light field. Only
pairwise interactions among excitons, evaluated in the Hartree-Fock (HF) approximation, and Pauli blocking modifications of the light-exciton coupling are included in the
nonlinear exciton dynamics. The theory is discussed in more detail in Section \ref{theory.sec}
below.
To facilitate conceptual understanding, we treat only the simple case of a circularly polarized polariton field in a single quantum well cavity [Fig.~\ref{sketch.fig}]. Owing to the spin dependence of the exciton-exciton interaction, \cite{takayama-etal.02} parametric scattering for linearly polarized polaritons is different from that in the circularly polarized case.\cite{schumacher-etal.07prb} There are indications \cite{Romanelli2007,schumacher-etal.07prb,ardizzone-etal.13} that transverse instabilities are more favored in the linearly polarized case. Moreover, using a double cavity also makes pattern formation easier.\cite{ardizzone-etal.13} We believe the conclusions drawn in this paper about physical mechanisms also apply to these more complicated cases.
We solve the equations of our theory in various
model state spaces. At the most complete level, we solve the
real($\bf r$)-space version of our equations directly on a grid.
While such simulations are realistic, they are computationally
demanding and it is not easy to disentangle the roles played by
various dynamical processes in generating the results. For this
latter purpose, we also evaluate the theory within reduced model
state spaces consisting of small numbers of $\bf k$ modes, the
relative simplicity of which makes them much more amenable to
analysis.
In our parameter-dependency study and our analysis of the
contributing dynamical processes, we make heavy use of a
single-hexagon model which has a state space consisting of one pump
mode (${\bf k} = {\bf 0}$) and six off-axis $\bf k$ modes arranged
in a regular hexagon. Larger models are also used to analyze
competitions among hexagons and more extended patterns.

This paragraph and the next summarize our findings. For
typical GaAs parameters, even if the system has complete azimuthal
isotropy, a ring pattern in momentum space is destabilized in favor
of a single hexagon with arbitrary orientation.
A small anisotropy is sufficient to destabilize the hexagon in favor of a two-spot pattern. As one increases the anisotropy continuously from zero, the transition from hexagon to two-spot is abrupt, and below this transition, a range of anisotropy values exist where both patterns are stable.
In the switching scheme, a two-spot pattern stabilized by anisotropy is set up as the initial signal, and the control beam is directed at another mode pair (see Fig.
\ref{switch.fig}). For our chosen material (GaAs) and pump beam parameters, the pattern can be switched by a control beam of intensity 1/6500 times the intensity of the two-spot signal.
In a detailed parameter variation study, when the control intensity is increased from zero, the stable state transits from the two-spot pattern
favored by the anisotropy, through an asymmetric hexagon (i.e. a
hexagon with asymmetric distribution of intensity), to the two-spot
pattern favored by the control. The transitions are again abrupt, with a bistable region at
each transition. The switching time scale is normally of the order of 1 ns but diverges as the control
intensity approaches from above the higher critical value for transition.

The analysis of the mechanisms underlying these phenomena has
yielded the following picture. The dynamics of the polariton
amplitude in a particular (off-axis $\bf k$) mode is driven by
pairwise scatterings for which the mode in question is one of the
outgoing modes. We classify the processes as linear, quadratic, or
cubic according to how many of the other three modes involved in the
scattering are off-axis (Fig. \ref{picofterms.fig}). The linear gain
processes provide the basic instability mechanism to the
pump-excited uniform polariton field.
For a pump intensity slightly above threhold, the set of $\bf k$ modes
experiencing linear instability growth form a ring of finite
thickness around the origin in (in-plane) momentum space. As the
polariton amplitudes grow in these modes, the quadratic and cubic
scattering processes, which represent interactions among the
off-axis modes, come into play. They drive the competitions among
phase-conjugate mode pairs (pairs with equal but opposite momenta
($\bf k$ and $- {\bf k}$)) which in turn determines whether a
particular pattern (ring, hexagon, two-spot etc) is stable. Because
of momentum conservation, the quadratic scatterings take place only
among off-axis modes that are on a regular hexagon (see Fig.
\ref{picofterms.fig} below), whereas the cubic processes are not
subjected to this restriction. Therefore, competitions among modes
residing on different hexagons are driven only by the cubic
processes. It is important to note that the scatterings occur among
coherent amplitudes: the relative phases of the involved polariton
amplitudes play a crucial role in determining whether a particular
process contributes to a gain or a loss in density for the mode in
question. Because of the complex feedback between the dynamics of
the phase and the magnitude of the polariton field, it is difficult
to cleanly disentangle the gain/loss effects of the different
processes.  Even so,
from the simulation results one can still characterize their effects
using concepts common in classical pattern forming systems as
follows. For the typical GaAs parameters used here, the stability of
the hexagonal pattern over two-spot patterns (in the isotropic case)
indicates that the net effect of the quadratic processes is
`cross-activating', i.e. the presence of density in one conjugate
mode pair favors the increase of density in other mode pairs. The
main effect of the cubic processes is to bring about saturation of
the off-axis polariton density. One can subdivide them into
`self-saturating' (the density in a conjugate mode pair inhibiting
its own growth) and `cross-saturating' (the density in one mode pair
inhibiting the growth of density in other pairs) processes. The
spontaneous breaking of the azimuthal symmetry of the system,
favoring the radially localized hexagons over a ring pattern,
indicates that the cross-saturating processes are stronger than the
self-saturating ones.

The above findings are based mostly on analyses of
models in which the pump intensity is infinitely localized in
momentum space (in the mode ${\bf k} = {\bf 0}$), or equivalently,
uniformly distributed over the 2D real space of the cavity's plane,
and the state space is truncated. Further investigations into the
effects of a finite pump laser spot size and inclusion of more modes
in the state space will be carried out within the full 2D real-space
simulation model.



The paper is organized as follows: We first describe and discuss our
theoretical approach in Section~\ref{theory.sec}. In Section \ref{models.sec}, we show a set
of results from a 2D $\bf r$-space simulation and discuss two
reduced models in truncated sets of $\bf k$ modes. In
Section~\ref{sinhex.sec}, we introduce and validate the
single-hexagon model through comparison with the previous models. In
Section \ref{parameter.sec}, we present the effects of variations of
anisotropy and control beam intensity on pattern selection and
switching behavior in the single-hexagon model. In Section
\ref{analysis.sec}, we classify the scattering/four wave mixing
processes among the polariton modes, and the complex interplay of
these processes are analyzed in Section \ref{interplay.sec}. An
analysis of the competitions among modes belonging to different
hexagons is carried out in Section \ref{no-quad.sec}. Section
\ref{summary.sec} contains some concluding comments.

\section{Theory of nonlinear optical response of a quantum
well microcavity}\label{theory.sec}

This section outlines the microscopic theory of the third-order
nonlinear response of a quantum well microcavity, from which our
working equations are derived. The theory can conceptually be
divided into two parts: the nonlinear response of the quantum well
to the presence of cavity photons and the coupling between the
relevant cavity photon modes and the light field outside of the
cavity. Since we work in the regime of strong linear coupling
between the QW exciton and a cavity photon mode, it is beneficial to
consider the polaritons as elementary excitations of the QW
microcavity and the cavity's nonlinear response as being due to the
interactions among the polaritons. For the theory of nonlinear QW
optical response, a more detailed account can be found in e.g. Refs.
\onlinecite{schaefer-wegener.02,kwong-binder.00,kwong-etal.01prb,takayama-etal.02}.
An account of our present theory was given in Ref.
\onlinecite{Dawes2009}.

In Ref. \onlinecite{Dawes2009}, our treatment of the passage of
light into and out of the cavity, when a given polaritonic
excitation is present inside the cavity, was only very briefly
explained. We start this section with a fuller account of this part
of our theory. Numerically, the most accurate way to propagate light
waves across the microcavity is, of course, to solve the Maxwell
equations throughout the DBR+QW structure using e.g. a transfer
matrix method
\cite{kwong-etal.01prb,yang-etal.05josab,stroucken-etal.96} or a
finite-difference time-domain method.\cite{kim-etal.10} However,
since we are concerned with obtaining an overall picture of
parametric dependencies rather than simulating a specific
experimental setup, we have chosen to use a simple model that
approximates the wave propagation effects as a coupling between the
radiation field outside of the cavity and the chosen cavity mode. We
will call the former the `macroscopic field'. In this model, the
whole microcavity is treated as a zero-thickness structure which
confines both the cavity photon field and the exciton field in the
longitudinal direction and lets them extend across the entire
transverse plane.  The effect of the cavity photon field on the
macroscopic light field is represented as a `polarization density'
in the wave equation governing the latter. Explicitly, we write the
electric displacement of the macroscopic field (which includes the
incident, reflected, and transmitted waves) as
\begin{equation}\label{D_ext.equ}
{\bf {\cal D}} ({\bf r} , t) = \epsilon_0 n^2_s {\bf {\cal E}} ({\bf
r} , t) - \hbar t_c {\bf E}_{\rm cav} (x,y,t) \delta(z)
\end{equation}
where ${\bf {\cal E}} ({\bf r} , t)$ is the macroscopic electric
field, $\epsilon_0$ is the vacuum permittivity, $n_s$ is the
refractive index of the substrate outside of the cavity, ${\bf
E}_{\rm cav} (x,y,t)$ is the cavity photon field, and $t_c$ is a
coupling strength. For later convenience, the dimension of $\hbar
t_c$ is chosen to be that of electric charge. In Eq.
(\ref{D_ext.equ}) we have set up a coordinate system in which the
$z$-axis is along the longitudinal direction and the cavity is
positioned at $z = 0$.

The incident waves (pump and control) come in from the left and are
all `$+$'-circularly polarized. The exciton spin states in the QW
are quantized along the $z$-axis so that an obliquely incident
($+$)-polarized beam may also generate excitons with spin$= - 1$
which, after interacting with other excitons, may generate a
($-$)-polarized component in the outgoing waves. However, the polar
angles of the off-axis beams are usually quite small, and we, for
simplicity, ignore this complication and take all outgoing fields to
be ($+$)-polarized. The algebraic development within the model is
detailed in Appendix \ref{field-coupling.sec}, with the following
results. Under the assumption that each of the waves involved is a
slowly varying envelope modulating a plane wave with a common
carrier frequency $\omega_p$, the electric field in the half-space
$z < 0$ can be written in the form
\begin{equation}\label{E_inc_refl_p.equ}
{\bf {\cal E}} ({\bf r}, t) = \sum_{\bf k} e^{i (k_x x + k_y y)}
{\hat {\bf e}}_+ \left[ {E}_{\mathbf{k},\text{inc}} (t + k_z (k,\omega_p)
z / \omega_p) - {E}_{\mathbf{k},\text{refl}} (t - k_z (k,\omega_p) z
/ \omega_p)\right] , \, z < 0
\end{equation}
where ${\bf k} = (k_x, k_y)$, $k = | {\bf k} |$, and $k_z(k,
\omega_p) = + \sqrt { {\omega^2_p n^2_s} / {c^2} - k^2}$, $c$
is the light speed in vacuum, and ${\hat {\bf e}}_+$ is the unit helicity vector for the ($+$)-polarized field. For each transverse momentum $\bf
k$, ${E}_{\mathbf{k},\text{inc}}$ is the incident field and
${E}_{\mathbf{k},\text{refl}}$ the reflected field.  In the $z > 0$
half plane, the field can be written as
\begin{equation}\label{E_trans_p.equ}
{\bf {\cal E}} ({\bf r}, t) = \sum_{\bf k} e^{i (k_x x + k_y y)}
{\hat {\bf e}}_+ {E}_{\mathbf{k},\text{trans}} (t - k_z (k,\omega_p) z /
\omega_p), \,\, z > 0
\end{equation}
where ${E}_{\mathbf{k},\text{trans}}$ is the transmitted field. The
cavity field, which by our argument above is also ($+$)-polarized,
is also decomposed into its spatial Fourier modes:
\begin{equation}\label{E_cav_p.equ}
{\bf E}_{\rm cav} (x, y, t) = \sum_{\bf k} e^{i (k_x x + k_y y)}
{\hat {\bf e}}_+ {E}_{\mathbf{k}} (t)
\end{equation}
We stress that ${E}_{\mathbf{k}} (t)$ here denotes a momentum-space
component of the cavity field and is not to be confused with the
macroscopic field ${\bf {\cal E}} ({\bf r}, t)$. The $\bf k$ sums in
Eqs. (\ref{E_inc_refl_p.equ})-(\ref{E_cav_p.equ}) range in principle
over the whole (transverse) momentum space, but they are restricted
to a finite set of $\bf k$ modes in the reduced-mode model
calculations discussed in this paper. For each $\bf k$ mode, the
transmitted (${E}_{\mathbf{k},\text{trans}}$), and reflected
(${E}_{\mathbf{k},\text{refl}}$) light fields are solved in terms of
the incident field (${E}_{\mathbf{k},\text{inc}}$) and the cavity
field (${E}_{\mathbf{k}}$) in Appendix \ref{field-coupling.sec}. The
result is:
\begin{eqnarray}
\label{Eeff.equ} {{E}_{\mathbf{k},\text{trans}}} &=&
{{E}_{\mathbf{k},\text{inc}}}-{{E}_{\mathbf{k},\text{refl}}}
\\\label{Erefl.equ}
{{E}_{\mathbf{k},\text{refl}}}&=&-\frac{\hbar
{{t}_{c}}}{2{{n}_{s}}c{{\epsilon
}_{0}}}\frac{d{{E}_{\mathbf{k}}}}{dt}
\end{eqnarray}

We next postulate the equation of motion of the cavity field under
the actions of the macroscopic light field and the exciton field. In
writing down the equation, we make the reasonable assumption that
each transverse $\bf k$ mode of the cavity field,
${{E}_{\mathbf{k}}}$, is a simple oscillator driven linearly by the
other two fields. The proportionality coefficients in the coupling
terms are chosen in such a way that $|{{E}_{\mathbf{k}}}|^2$ can be
interpreted as mode photon density (see Appendix
\ref{field-coupling.sec}). The equation is:
\begin{equation}\label{Ek.equ}
i\hbar \frac{d{{E}_{\mathbf{k}}}}{dt}=\hbar \omega
_{\mathbf{k}}^{c}{{E}_{\mathbf{k}}}-{{\Omega
}_{\mathbf{k}}}{{p}_{\mathbf{k}}}+\hbar
{{t}_{c}}E_{\mathbf{k},\text{inc}}^{\text{eff}} ,
\end{equation}
where $\omega _{\mathbf{k}}^{c}$ is the uncoupled mode dispersion
relation for a cylindrically symmetric cavity: $\omega
_{\mathbf{k}}^{c} = \sqrt{{\omega_{\mathbf{0}}^{c}}^2 + c^2 k^2 /
n^2_b}$, $\omega_{\mathbf{0}}^{c}$ being the cavity mode energy at
${\bf k} = {\bf 0}$, and $n_b$ being an effective refractive index
of the cavity's medium which characterizes the transverse
propagation speed of the cavity mode. ${p}_{\mathbf{k}}$ denotes the
exciton field, which will be identified below with the $1s$
heavy-hole exciton contribution to the interband polarization in the
QW, so that $| {p}_{\mathbf{k}} |^2$ is the exciton density (number
of excitons per unit area) in the $\bf k$ mode.{\footnote {Strictly
speaking, $| {p}_{\mathbf{k}} |^2$ and $|{{E}_{\mathbf{k}}}|^2$ only
represent the coherent parts of their respective densities. But
since the incoherent densities are expected to be small and are
ignored in this paper, we will drop the qualifier `coherent'.} The
exciton-photon coupling strength ${\Omega }_{\mathbf{k}}$ is an
input parameter in Eq. (\ref{Ek.equ}) but will be given a
microscopic meaning in terms of electron-hole dynamics below.
$E_{\mathbf{k},\text{inc}}^{\text{eff}}$ is the (limiting value of
the) macroscopic electric field at the cavity's position. Within our
model, we have
\begin{equation}\label{E_eff_trans.equ}
E_{\mathbf{k},\text{inc}}^{\text{eff}} (t) =
{E}_{\mathbf{k},\text{trans}} (t)
\end{equation}
The coupling coefficient $t_c$ was introduced in Eq.
(\ref{D_ext.equ}) above. If we set the incident field
${E}_{\mathbf{k},\text{inc}}$ and the exciton field
${p}_{\mathbf{k}}$ to zero, Eqs.
(\ref{Eeff.equ})-(\ref{E_eff_trans.equ}) would reduce to a
homogeneous equation for ${E}_{\mathbf{k}}$ from which we can
extract a radiative decay rate (in energy units) $\Gamma = \omega_p \hbar^2
t_c^2/(2\epsilon_0 c n_s)$ for the cavity.  We reiterate that Eq.
(\ref{Ek.equ}) is written down as a {\em postulate} based on
physically reasonable assumptions. This treatment is sufficient for
our purposes here. Alternatively, it could be derived as a classical
approximation in a theory that starts with the cavity photon field
being quantized.

We next briefly describe the microscopic theory of the excitonic
optical response of quantum wells which gives the equation of motion
of ${p}_{\mathbf{k}}$.
\cite{schaefer-wegener.02,kwong-binder.00,kwong-etal.01prb,takayama-etal.02}
It starts with a many-particle Hamiltonian operating on the Fock
space of electrons and holes. Referring to Fig.~\ref{sketch.fig}
again, for the quantum well, we consider a GaAs-type semiconductor
band structure around the fundamental band gap with an exciton size
of about $10^{-6} {\rm ~cm}$. For the present study, the
single-particle basis includes only one parabolic conduction band
and one parabolic valence band, i.e. the heavy hole band. The
Hamiltonian contains the charge carriers' kinetic energy and
pairwise Coulomb interactions among them, and a coupling to the
cavity photon field that either creates or annihilates an
electron-hole pair in the rotating wave approximation. Thus the only
material parameters in the theory are the electron mass, the hole
mass, the background dielectric constant, and the interband dipole
coupling. With this Hamiltonian, equations of motion of observables
(certain expectation values of products of charge carrier
creation/annihilation operators) are derived.  The dynamical
variable relevant to us, the exciton field $p_{\bf k} (t)$, is
defined in this microscopic theory as the interband polarization
restricted to the heavy-hole $1s$ exciton subspace:\cite{takayama-etal.02}
\begin{equation}
\label{p-def.equ} p_{\bf k} (t) = \frac{1} {{\cal L}^2} \sum_{
{\bf k}' } \tilde{ \phi } ( {\bf k}' + \beta {\bf k} ) \langle
a_{h,-{\bf k}'} (t) a_{e,{\bf k}'+{\bf k}} (t) \rangle
\end{equation}
where $a_{e,{\bf q}} (t)$ is the Heisenberg-picture annihilation
operator of a conduction band electron with momentum $\bf q$ at time
$t$ and $a_{h,{\bf q}} (t)$ is the corresponding operator for a hole
state, and $\langle \cdot \cdot \cdot \rangle$ represents the
expectation value in the initial state of the electron-hole system.
$\beta=m_h/(m_e+m_h)$ is the ratio of the hole mass to the total
exciton mass, ${\cal L}^2$ is the area of the normalization box, and
$\tilde{ \phi} ( {\bf q}) = {\sqrt{2 \pi} a_0} / {[1+(a_0 q / 2 )^2
]^{3/2}}$ is the two-dimensional internal $1s$ exciton wavefunction
in the space of relative electron-hole momentum, with $a_0 = \hbar^2
\epsilon_b/(e^2 m_r)$, $\epsilon_b$ and $m_r$ being the (3D) exciton
Bohr radius, the background dielectric constant and the reduced mass
of the electron-hole pair respectively. The restriction to the $1s$
state is valid because we limit the pump intensity so that the
maximum generated polariton density is of the order of $10^{10}
{\rm~cm}^{-2}$. At this density and the chosen pump frequency, which
is tuned to a frequency within the lower polariton branch
(Fig.~\ref{sketch.fig}(b)), the excitations in the quantum well are
expected to stay predominantly in the $1s$ heavy hole exciton state.
For a more detailed analysis on the validity of the $1s$
approximation, see Ref.~\onlinecite{takayama-etal.02}.
Since we restrict our considerations to $+$-polarized excitons, the
electron and the hole in Eq. (\ref{p-def.equ}) have spins $-1/2$ and
$3/2$ respectively.

Starting from the electron-hole vacuum as the initial state, the
equation of motion for the $p_{\bf k} (t)$ is derived up to third
order in the light field amplitude \cite{kwong-binder.00,
axt-stahl.94, schaefer-wegener.02} on the Hartree-Fock-level. In the
derivation, the single-pair electron-hole basis is restricted to the
$1s$ state as mentioned. In this limit, the equation for a system of
co-circularly-polarized excitons in the coherent third-order regime
is:\cite{kwong-etal.01prb,takayama-etal.02,schumacher-etal.09}
\begin{equation}\label{pk.equ}
i\hbar\frac{d{{p}_{\mathbf{k}}}}{dt} = \left(
\varepsilon_{\mathbf{k}}^{x} -i{\gamma_x}\right)p_{\mathbf{k}} -
\Omega_{\mathbf{k}} E_{\mathbf{k}} + \frac {1} {{\cal L}^2}
\sum\limits_{\mathbf{q}\mathbf{k'}\mathbf{k''}} \left(2\tilde{A}
\Omega_{\mathbf{k''}}p_{\mathbf{q}}^{*} p_{\mathbf{k'}}
E_{\mathbf{k''}} + V_{\text{HF}} p_{\mathbf{q}}^{*} p_{\mathbf{k'}}
p_{\mathbf{k''}}\right)\delta_{\mathbf{q},\mathbf{k'}+
\mathbf{k''}-\mathbf{k}}
\end{equation}
Here $\varepsilon_{\mathbf{k}}^{x}$ is the free exciton energy and
$\gamma_x$ is a phenomenological dephasing rate which represents all
non-radiative losses, including the transfer of coherent excitons to
the incoherent population. The photon-exciton coupling
$\Omega_{\mathbf{k}}$, which is an input parameter in the cavity
photon equation, Eq. (\ref{Ek.equ}), is given in the present
microscopic theory by (in S.I. units)
\begin{equation}\label{om.eq}
\Omega_{\mathbf{k}} = e \langle r \rangle_{e h} \phi^*({\bf 0})
f_{\gamma} (z_{\rm QW}) \sqrt{\hbar \omega _{\mathbf{k}}^{c} /
\epsilon_0 \epsilon_b},
\end{equation}
where $\langle r \rangle_{e h}$ is the interband dipole matrix
element, $\phi^*({\bf 0})=2\sqrt{2}/(a_0 \sqrt{\pi})$ is the
two-dimensional real space $1s$ exciton wavefunction (Fourier
transform of $\tilde{\phi} ( {\bf q})$) at zero electron-hole
separation, $f_{\gamma}(z)$ is the resonant cavity photon
wavefunction along the coordinate axis ($z$-axis) normal to the
cavity's plane, normalized to unity: $\int^{\infty}_{- \infty} dz |
f_{\gamma} (z) |^2 = 1$ and $z_{\rm QW}$ is the position of the
quantum well on the $z$-axis.

Of the two third-order terms, the term with the $pp^*E$ structure
stems from Pauli blocking, or phase-space filling (PSF), among the
fermionic components of the exciton that are created by photon
absorption and those of excitons already in the system. The strength
is given by $\tilde{A} = 2 \pi a^2_0 /7$. The other term is due to
mean-field, or Hartree-Fock (HF), Coulomb interaction among the
excitons, with strength $V_{\text{HF}} = 2 \pi E_b a_0^2 (1-315
\pi^2 /4096) \approx 1.52 E_b a_0^2$, where $E_b$ is the (2D)
exciton Rydberg energy. Both $\tilde{A}$ and $V_{\text{HF}}$
actually depend on the momenta of the involved excitons and photons,
but since the dependencies are rather weak and the relevant momenta
are not excessively large, we approximate them by their values at
zero momentum. Each exciton interaction or PSF term can be
visualized as either a four-wave mixing process with wave-vector and
frequency matching or a polariton-polariton scattering process with
momentum and energy conservation. The full set of third-order terms
also includes a time-retarded Coulomb correlation between two
excitons, which is neglected here. Because of its time-nonlocal
structure, including the Coulomb correlation term in
Eq.~(\ref{pk.equ}) would increase the complexity in solving the
equation substantially. It was previously shown
\cite{kwong-etal.01prb,takayama-etal.02} that it is much less
important than the PSF and HF terms when the two interacting
excitons are co-circularly polarized and have energies far below the
exciton resonance. So dropping the retarded-Coulomb term in our case
is expected to be a good approximation. A detailed study of its
effects on polariton pattern formation will be a subject for future
work.

Moreover the dynamical effects of the dephased excitons are not
considered. A more comprehensive theory would include a description
of the transfer by dephasing and the subsequent interactions between
the coherent and incoherent populations. Equations of motion serving
this purpose could be derived within the microscopic frameworks that
produce Eq.~(\ref{pk.equ}). While recent experiments indicate that
the incoherent exciton population remain small
\cite{ballarini-etal.09} in the density and frequency regions that
we consider, its cumulative effects should be studied more
carefully. Similar considerations apply to three-exciton and
larger-cluster correlations which would be accounted for in an
extended theory beyond the $\chi^{(3)}$ regime.

To recapitulate, Eqs. (\ref{Eeff.equ}), (\ref{Erefl.equ}), (\ref{Ek.equ}),
(\ref{E_eff_trans.equ}), and (\ref{pk.equ}) are the working
equations in the present paper.  With the incident
macroscopic light field ${E}_{\mathbf{k},\text{inc}} (t)$ as input, the
response of the quantum well microcavity, $E_{\mathbf{k}} (t)$ and
${p}_{\mathbf{k}} (t)$, as well as the resulting output light
fields, ${{E}_{\mathbf{k},\text{trans}}} (t)$ and
${E}_{\mathbf{k},\text{refl}} (t)$, are calculated by solving this
set of equations simultaneously.



\section{Simulations and reduced models}
\label{models.sec} In this section, we show some representative
numerical results on transverse pattern competition and control in a
quantum well microcavity by solving, at various levels of
approximation, Eqs. (\ref{Eeff.equ})-(\ref{E_eff_trans.equ}), and
(\ref{pk.equ}) laid down in the previous section. The numerical
challenge stems mainly from the nonlinear (HF and PSF) terms in the
exciton field equation, Eq. (\ref{pk.equ}), which couple the $\bf k$
modes together. Since the planar normalization box (in real space)
is taken to infinity, the summation sign in Eq. (\ref{pk.equ})
actually denotes $\bf k$ space integrations. A comprehensive
simulation of the dynamics of the quantum-well microcavity then
involves solving the equations on a two-dimensional momentum space
grid, or, alternatively, solving the Fourier transforms of the
equations on a real space grid in the plane of the microcavity.
While such simulations are realistic, they are computationally
demanding, and their results are not easy to analyze. Since, when
the system is not far above the instability threshold, the
stationary transverse patterns are usually distributed over
localized regions in $\bf k$ space, approximations in which the
equations are solved within judiciously chosen, vastly reduced sets
of $\bf k$ modes can profitably be used. In this paper, we take
advantage of this fact and put our emphasis on dissecting the
results from several such reduced models in order to advance towards
our aim of understanding the physical mechanisms of pattern
competition in our system.

In the following, after some preliminary remarks on the linear
instability of the spatially planar homogeneous (with only the ${\bf
k} = {\bf 0}$ component) state, we will present a set of
representative results of our full two-dimensional simulations.  We
will then introduce two reduced models -- the `multi-$| {\bf k} |$
model' and the `ring model' -- in which the equations of motion are
solved over selected sets of $\bf k$ modes. These two models
essentially reproduce the representative results of the full 2D
simulations. We will extend our investigations using these models.
Based on these investigations, we will argue at the end of this
section that the analysis of an even simpler model, the state space
of which consists of the ${\bf k} = {\bf 0}$ mode and a hexagonal
set of six off-axis modes, could be expected to provide the critical
insights to understand the patterns' behaviors in the more
comprehensive models. This `single-hexagon' model will be analyzed
in detail in the following sections.

For the calculations, material parameters appropriate
for GaAs are used: $E_b \approx 13~{\rm meV}$, $a_0 \approx 170~{\rm
\AA}$, $\varepsilon_{\bf 0}^x = 1.497\, \text{eV}$, $\gamma_x =
0.4\, \text{meV}$, $\Omega_k \equiv \Omega = 8\, \text{meV}$, $n_b =
3.6$, and $\epsilon_b = n^2_b$. With the given values of $E_b$ and
$a_0$, ${V}_{\text{HF}} \approx 0.57 \times 10^{-10} \text{meV}
\text{cm}^{2}$, and $2\tilde{A}\Omega$ is about three times weaker.
For the cavity mode, we choose $\hbar \omega_{\bf 0}^c =
\varepsilon_{\bf 0}^x$, and $t_c$ is chosen such that the radiative
decay rate is $\Gamma \approx 1.5\, \text{meV}$ for $\hbar \omega_p$
around $1.5 \text{eV}$. The incident pump is tuned to $5~\text{meV}$
below the bare exciton resonance, i.e. $\hbar \omega_p = 1.492
\text{eV}$, and its steady state flux intensity, which we denote by
$I_{\text{pump}}$, is about $25.2~\text{kW} \text{cm}^{-2}$.

\begin{figure}[b]
\includegraphics[scale=0.2,angle=0,trim=00 00 00
00]{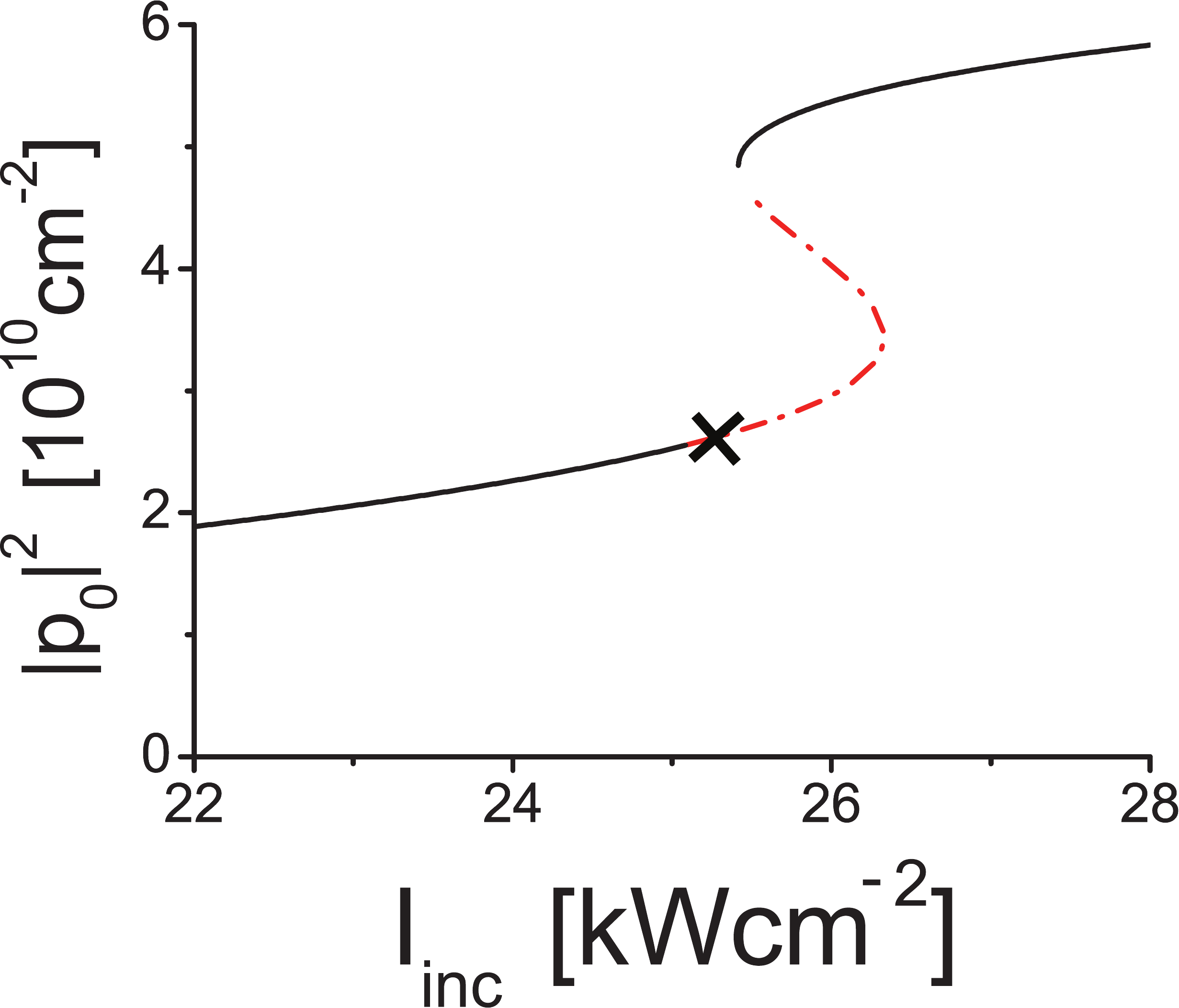} \caption{(Color online.) Steady state pump-generated
exciton density in the quantum well as a function of the pump beam
intensity. For densities inside the red, dashed segment, the uniform
polariton field is unstable against finite $| {\bf k} |$
perturbations. The cross marks the pump intensity used in the
simulations here.}\label{scurve.fig}
\end{figure}

The linearized Eq.~(\ref{Ek.equ}) and Eq.~(\ref{pk.equ}) at the
lossless limit, i.e. with ${V}_{\text{HF}}$, $\tilde{A}$,
$\gamma_x$, and $t_c$ set to zero, give the linear polariton
spectrum and the polariton eigenvectors in the photon-exciton basis.
With the other parameters set as above, the linear polariton
spectrum is plotted in Fig. \ref{sketch.fig}.  Returning to the full
nonlinear equations, one can easily verify that when the input field
is incident normally on the cavity (pump only), there exist
solution(s) that preserve the planar translational symmetry of the
input beam ($p_{\bf k}$ and $E_{\bf k}$ vanish for all ${\bf k} \neq
{\bf 0}$). At steady state, $p_{\bf 0}$ and $E_{\bf 0}$ oscillate at
the pump frequency $\omega_p$. The exciton density $| p_{\bf 0} |^2$
calculated for the above parameter values is plotted as a function
of the incident field intensity $I_{\text{inc}} = |
{E}_{\mathbf{0},\text{inc}} |^2$ in Fig.~\ref{scurve.fig}.
A linear stability analysis with off-axis (${\bf k} \neq {\bf 0}$)
perturbations on the uniform solution is performed, and the range of
instability against this class of perturbations is also shown in
Fig.~\ref{scurve.fig}. The stability matrix is block-diagonal in
$\bf k$ space with the fields $p_{\bf k}, E_{\bf k}, p_{- {\bf k}},$
and $E_{- {\bf k}}$ for each $\bf k$ coupled together. For each
value of $| p_{\bf 0} |$ within the off-axis instability range, the
uniform state is unstable against small random perturbations in a
set of $\bf k$ modes, which form a ring of finite radial thickness
in $\bf k$ space. The mean radius of this ring is approximately determined by phase matching that fulfills the resonance in the polariton scattering, as illustrated in Fig. \ref{sketch.fig}b. The dispersion relations in Fig. \ref{sketch.fig}b are calculated without the exciton interactions in Eq. (\ref{pk.equ}), which shift the exciton energy, and hence the (lower) polariton energy, upwards. As a result, the actual resonant $| {\bf k} |$ value is smaller than that shown in Fig. \ref{sketch.fig}b. The point ($| {E}_{\mathbf{0},\text{inc}} |$, $|
p_{\bf 0} |$) corresponding to the incident field strength used in
our calculations is marked in Fig. \ref{scurve.fig}. For this value,
the instability-triggering ${\bf k}$ modes lie in the range
$1.25\times 10^{-4} \text{cm}^{-1}$ $\leq| {\bf k} | \leq $
$1.40\times 10^{-4} \text{cm}^{-1}$. More detailed discussions of
the linear stability analysis of the uniform state may be found in
Refs. \onlinecite{luk.12,schumacher-etal.07prb}. In our calculations, starting from the
uniform state under steady pump irradiation, the polariton field
(the lower polariton branch) for each mode inside the `ring of
instability' initially grows in time exponentially from random
fluctuations. Subsequently, the interactions among these off-axis
polariton modes, and the ${\bf k} = {\bf 0}$ mode, govern the
long-time competition dynamics among them.

\subsection{2D real-space simulations} \label{full-2D.sec}

The full 2D simulations are performed in real space by solving the
Fourier transforms of Eqs.~(\ref{Ek.equ}) and (\ref{pk.equ}). We
show a representative set of results here. A more comprehensive
discussion of these simulations will be presented in a future
publication. The configuration space is a square (spatial) grid with
a step size of $\Delta x = \Delta y = 0.45 \mu{\rm m}$ in a box of
length $ 90 \mu{\rm m}$ on one side. The initial state of the system
is the polariton `vacuum', i.e., the ground state of the
semiconductor microcavity. The pump beam is switched on at $t=0$,
its intensity becoming steady after 8 ns. Shown in
Fig.~\ref{full2d.fig}(a), the steady-state pump profile is broad and
flat so that its corresponding (transverse) momentum distribution is
narrowly peaked at ${\bf k} = {\bf 0}$. As time progresses, coherent
polariton amplitudes in several off-axis $\bf k$ modes form ``spontaneously",
causing density patterning inside the cavity. The
calculated real-space exciton density distribution inside the QW at
$t = 18 {\rm ns}$ is shown in Fig.~\ref{full2d.fig}(b). The
corresponding momentum distribution is plotted in
Figure~\ref{full2d.fig}(c), which shows a dominant hexagonal pattern
situated within the ring of linear instability. The azimuthal
orientation of the hexagon is random.

\begin{figure}
\includegraphics[scale=0.3, angle=0, trim = 2cm 1.8cm 2.5cm 0cm]{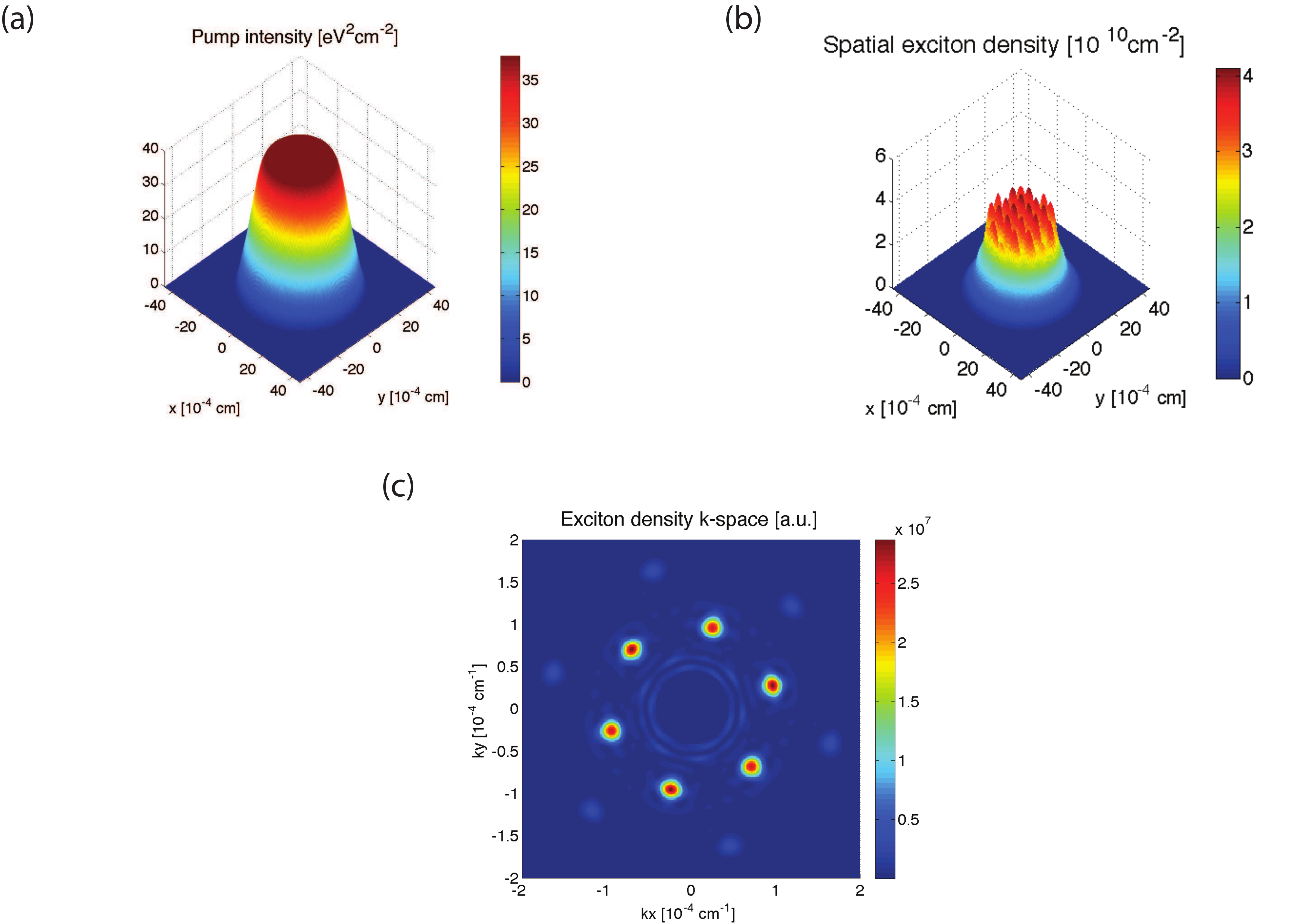}
\caption{ (Color online.) Two-dimensional real-space simulations of
hexagonal pattern formation. (a) The pump profile in 2-dimensional
real space. (b) The exciton density profile in the QW with
modulations stemming from transverse instabilities. (c) Hexagonal
pattern in transverse momentum space formed by the exciton field in
(b).  The density in the center (in the pump mode) has been masked
for clarity. See text for details of the simulations.}
\label{full2d.fig}
\end{figure}




\subsection{Reduced models}\label{red-models.sec}

The linear off-axis instability spontaneously breaks the
(transverse) translational symmetry of the system's setup but
preserves the latter's azimuthal symmetry. In the full 2D
simulations, when the off-axis polariton fields are being built up,
their mutual interactions also break the azimuthal symmetry,
resulting in a stable hexagonal structure. As explained above, the
simulation results indicate that the collection of modes taking part
in the pattern dynamics are relatively localized in $\bf k$ space.
We take advantage of this fact and use reduced models in the
following to investigate the competitions among different patterns,
their control, and the physical mechanisms underlying these
competitions.

In the reduced models, we restrict our state space to a selected
finite set of $\bf k$ modes. We introduce two such models in this
subsection: (i) The `multi-$| {\bf k} |$ model'. For the state space
in this model, we choose six directions corresponding to the
vertices of a regular hexagon centered at ${\bf k} = {\bf 0}$. The
hexagon's orientation is arbitrary. $N$ evenly spaced points along a
radial segment in each direction are included in the state space, as
illustrated in Fig.~\ref{models.fig}(a). Explicitly, we write the
$\bf k$ points as:
\begin{equation}\label{multi-k-pt.equ}
{\mathbf{k}_{h,i}}=(k_0+h \delta k) \hat{\bf e}_i \,\, , \,\,
h=1,\cdots, N \,\, , \,\, i=1,\cdots, 6
\end{equation}
where $\hat{\bf e}_i,i=1,\cdots, 6$ are unit vectors in the vertex
directions, and the minimum radius $k_0$ and the grid size $\delta
k$ are run-time parameters in the numerical calculations. (ii) The
`ring model'. In this model, we choose the $|{\bf k}|$ modes to lie
on a circle centered at the origin, regularly spaced in the angle
coordinate, as illustrated in Fig.~\ref{models.fig}(b). We also
organize the points into $N$ hexagons, writing them as
\begin{equation}\label{ring-k-pt.equ}
{\mathbf{k}_{h,i}}=k_r \hat{\bf e}_{h,i} \,\, , \,\, h=1,\cdots, N
\,\, , \,\, i=1,\cdots, 6
\end{equation}
where $\hat{\bf e}_{1,i}$ is a unit vector in an arbitrarily
selected direction, $\hat{\bf e}_{h+1,i}$ is obtained by rotating
$\hat{\bf e}_{h,i}$ counterclockwise through the angle $\frac {\pi}
{3 N}$, and the radius of the circle, $k_r$, is a run-time
parameter.

\begin{figure}
\includegraphics[scale=0.30, angle=0]{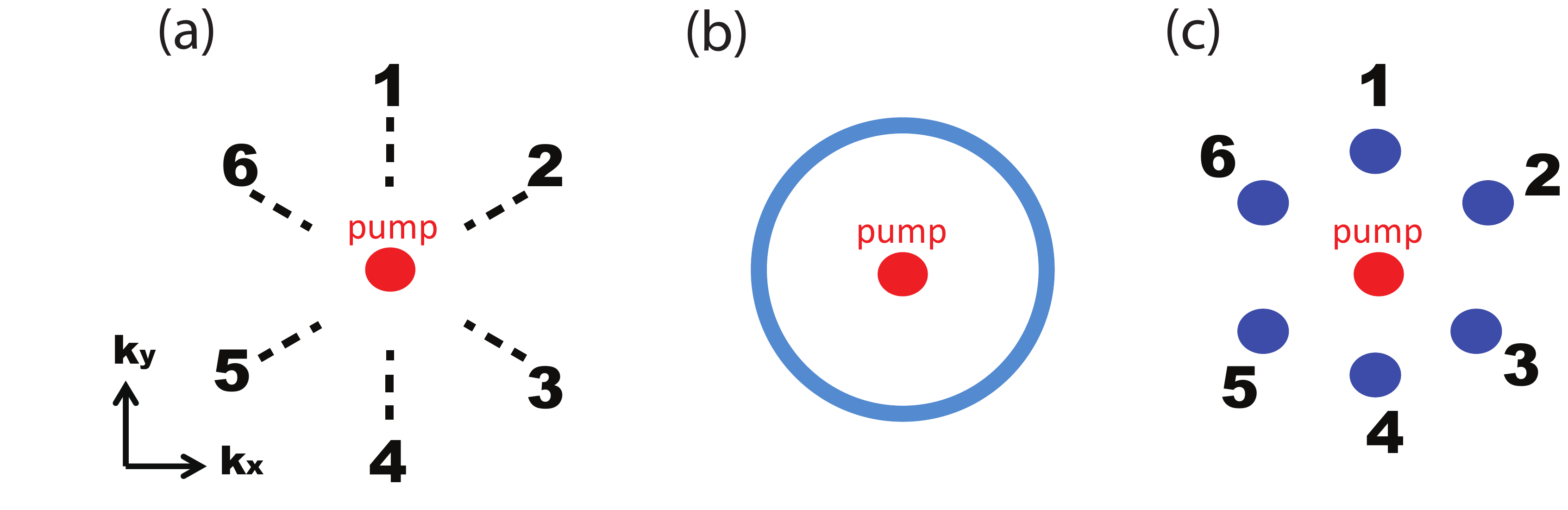}
\caption{(Color online.) Sketches of the state spaces, made up of
modes in transverse momentum space, of the three reduced models used
in this paper. (a) Multi-$|{\bf k}|$ model. Beside the origin, the
modes lie on six radial segments (thick dashed lines) arranged
hexagonally. (b) Ring model. The off-axis part of the state space is
a ring of linearly unstable $\bf k$ modes. (c) Single-hexagon model.
The state space is the origin plus a six-mode subset of the ring
model state space that form a regular hexagon. }\label{models.fig}
\end{figure}

By considering the momentum conservation constraint in the nonlinear
terms in Eq.~(\ref{pk.equ}), one can see that, for both models,
Eq.~(\ref{pk.equ}) reduces to equations of the following form (we
have simplified the notation $p_{\mathbf{k}_{h,i}}$ to $p^{}_{h,i}$
and $p_{\bf 0}$ to $p_0$). The $\mathbf{k}=0$ (on axis) equation
reads:

\begin{eqnarray} \label{P0.equ}
i\hbar {{\dot{p}}_{0}} & = & \left[ \varepsilon _{0}^{x}-i{{\gamma
}_{x}}+{{V}_{\text{HF}}}\left( {{\left| {{p}_{0}}
\right|}^{2}}+2\sum\limits_{h,i}{{{\left| {{p}_{h,i}}
\right|}^{2}}} \right)+2\tilde{A}\Omega
\sum\limits_{h,i}{p_{h,i}^{*}{{E}_{h,i}}} \right]{{p}_{0}}
\nonumber\\ && -\Omega \left[ 1-2\tilde{A}\left( {{\left|
{{p}_{0}} \right|}^{2}}+\sum\limits_{h,i}{{{\left| {{p}_{h,i}^{}}
\right|}^{2}}} \right) \right]{{E}_{0}}
+p_{0}^{*}\sum\limits_{h,i}\left( 2\tilde{A}\Omega
p_{h,i}^{}{{E}_{h,i+3}^{}}+{{V}_{\text{HF}}} p_{h,i}^{}
p_{h,i+3}^{} \right) \nonumber\\ && +
2\sum\limits_{h,i}{p_{h,i}^{*}}\left[ \tilde{A}\Omega \left(
p_{h,i+1}^{}{{E}_{h,i-1}^{}}+p_{h,i-1}^{}{{E}_{h,i+1}^{}}
\right)+{{V}_{\text{HF}}}p_{h,i+1}^{} p_{h,i-1}^{} \right] .
\end{eqnarray}

\noindent{The off-axis equations, for $h = 1, \cdots, N$ and $i = 1,
\cdots, 6$, read:}
\begin{eqnarray} \label{Pi.equ}
i\hbar {\dot{p}}_{h,i}^{} & = & \left[ \varepsilon
_{h,i}^{x}-i{{\gamma }_{x}}+2{{V}_{\text{HF}}}\left( {{\left|
{{p}_{0}} \right|}^{2}}-\frac{1}{2}{{\left| {{p}_{h,i}^{}}
\right|}^{2}}+\sum\limits_{h',i'}{{{\left| {{p}_{h',i'}^{}}
\right|}^{2}}} \right)+2\tilde{A}\Omega
\left(p_0^{*}{{E}_{0}}+\sum\limits_{h',i'}{p_{h',i'}^{*}{{E}_{h',i'}^{}}}
\right) \right]{{p}_{h,i}^{}} \nonumber \\&& - \Omega \left[
1-2\tilde{A}\left( {{\left| {{p}_{0}} \right|}^{2}}-{{\left|
{{p}_{h,i}^{}} \right|}^{2}}+\sum\limits_{h',i'}{{{\left|
{{p}_{h',i'}^{}} \right|}^{2}}} \right) \right]{{E}_{h,i}^{}}
\nonumber \\&& + 2p_{h,i+3}^{*}\left\{ \tilde{A}\Omega \left[
{{p}_{0}}{{E}_{0}}+\sum\limits_{h',i'}{p_{h',i'}^{}{{E}_{h',i'+3}^{}}-\left(
p_{h,i}^{}{{E}_{h,i+3}^{}}+p_{h,i-3}^{}{{E}_{h,i}^{}} \right)}
\right] \right.\nonumber \\&& \left. +{{V}_{\text{HF}}}\left[
\frac{1}{2}p_{0}^{2}+\sum\limits_{h'}{\left( p_{h',i+1}^{}
p_{h',i-2}^{}+p_{h',i+2}^{}p_{h',i-1}^{} \right)+\sum\limits_{h'\ne
h}{p_{h',i}^{}p_{h',i-3}^{}}} \right] \right\} \nonumber \\&&
+2p_{0}^{*}\left[ \tilde{A}\Omega \left(
p_{h,i+1}^{}{{E}_{h,i-1}^{}}+p_{h,i-1}^{}{{E}_{h,i+1}^{}}
\right)+{{V}_{\text{HF}}}p_{h,i+1}^{}p_{h,i-1}^{} \right] \nonumber
\\&& +2p_{0}\left[ \tilde{A}\Omega \left(
p_{h,i+2}^{*}{{E}_{h,i+1}^{}}+p_{h,i-2}^{*}{{E}_{h,i-1}^{}}
\right)+{{V}_{\text{HF}}}\left(
p_{h,i+2}^{*}p_{h,i+1}^{}+p_{h,i-2}^{*}p_{h,i-1}^{} \right) \right]
\nonumber \\&& +2\tilde{A}\Omega \left(
p_{h,i+2}^{*}p_{h,i+1}^{}+p_{h,i-2}^{*}p_{h,i-1}^{} \right){{E}_{0}}
\end{eqnarray}
When the off-axis direction number subscript $i$ is outside the
range $1-6$, it represents the direction either $i-6$ for $i$
greater than $6$ or direction $i+6$ for $i$ smaller than $1$.
For simplicity we have neglected the $\bf k$ dependence of
$\Omega_{\bf k}$ and replaced it by $\Omega = \Omega_{\bf 0}$. The
cavity field equations,
Eqs.~(\ref{Eeff.equ})-(\ref{E_eff_trans.equ}), remain the same for
each ${\mathbf{k}_{h,i}}$, with analogous notations being used for
the cavity and macroscopic fields: $E_{h,i}$ etc.

On the right hand side of Eq. (\ref{P0.equ}), the $\chi^{(3)}$ terms
represent, in the order of their appearance, the HF shift
in the exciton energy, photon scattering off an exciton density
grating, Pauli blocking of the photon-exciton coupling, and various
density transferring scattering processes.  The terms in Eq.
(\ref{Pi.equ}) can also be interpreted the same way. We will discuss
the density transferring processes in more details below.

\subsubsection{The multi-$|{\bf k}|$ model: results}
Within this model we explore the competitions among patterns that
exist in a hexagonal geometry and the radial distributions of these
patterns in $\bf k$ space. The state space consists of $N = 256$
$|{\bf k}|$ points in the direction of each of the six hexagonal
vertices. We use a grid size of $22.5~\text{cm}^{-1}$ along each
radial segment so that the state space is spread over the range from
$1.10675\times 10^4~\text{cm}^{-1}$ to $1.6805\times 10^4
~\text{cm}^{-1}$ in each segment. In this state space, Eqs.
(\ref{Eeff.equ})-(\ref{E_eff_trans.equ}), (\ref{P0.equ}), and (\ref{Pi.equ}) are solved in the rotating
frame of the incident field, i.e. with $e^{- i \omega_p t}$ factored
out of all the fields in the equations. The time-stepping is
performed with a 4-th order Runge-Kutta method with a time step of
$20~\text{fs}$. This time step is sufficient for simulating the
system dynamics, which has a typical time scale of the order of
$1~\text{ps}$, with good accuracy. Random fluctuating light
sources ($1.26 \times 10^{-6}~\text{W} \text{cm}^{-2}$) are added in the off-axis
modes as seeds for the initial instability growth.

Solving this model with the same material and beam parameters as in
the full-2D simulations yields similar results, i.e., starting with
the 2D uniform state, the system spontaneously generates off-axis
coherent fields which subsequently stabilize as a hexagonal pattern.
Along each radial segment, the off-axis field initially grows in
modes over a range of $| {\bf k} |$ values. When the pattern is
stabilizing, however, one mode in each hexagonal direction wins, the
intensity in the other modes in the segment decaying to zero.

In the experiment using atomic vapors in Ref.
\onlinecite{Dawes2005}, a stable two-spot pattern was obtained
instead of a hexagon. It was theorized that an imperfect azimuthal
symmetry, which is unavoidable in practice, might have favored the
two-spot pattern. In the theoretical calculations on microcavities
in Ref. \onlinecite{schumacher-etal.09}, it was found that a small
anisotropy introduced in one orientation is sufficient to favor the
two-spot over the more symmetric distribution over all six hexagonal
directions, and the subsequent controlled switching involves another
two-spot pattern. For a recent detailed discussion of anisotropy in
a real system, see Ref.~\onlinecite{Abbarchi2012}.

We have carried out
calculations for this slightly anisotropic situation. The anisotropy
is imposed, as in Ref. \onlinecite{schumacher-etal.09}, by lowering
$\omega_k^c$ by $0.12~\text{meV}$ in directions 1 and 4 at every
$|{\bf k}|$. The values of other parameters remain as set above. A set of
numerical results that include reversible switching by a control
beam are shown in Fig.~\ref{erefl.fig}. The top three panels show
the light intensities reflected from the microcavity in the six
off-axis directions, and the bottom panel shows the control beam, as
functions of time. In each direction, the intensity has been summed
over all the modes on the radial segment in
Fig.~\ref{models.fig}(a). The simulation starts at time $t = 0$ with
all fields equal to zero except for some low-level fluctuations. The
external pump field is turned on in the ${\bf k} = {\bf 0}$ mode and
reaches its steady state at about 30 ps with $|p_0|^2$ = $2.6 \times
10^{10}~{\rm cm}^{-2}$. The pump steady state intensity is just above
the instability threshold, leading to the build-up of field
intensity in the off-axis modes. The build-up occurs in all six
directions initially, but after a while, the asymmetry (anisotropy)
introduced in the cavity mode's dispersion relation gives a decisive
advantage to directions 1 and 4, where the field intensity continues
to grow to a steady state at around $t \approx 2.7~\text{ns}$, while
the intensities in the other four directions fall back to
fluctuation level. A control beam of intensity $= 9.1 \times 10^{-3}~\text{W} \text{cm}^{-2} = 3.6
\times 10^{-7} I_{\text{pump}}$ is introduced in direction 2 (see
below for the radial $| {\bf k} |$ value of the control) at $t =
3~\text{ns}$, and it switches the off-axis intensity to directions 2
and 5, reaching a steady state level after 1.6~ns. In the switching
process, the instability-generated field intensity is about 6500 times
stronger than the control intensity. The control beam is then turned
off at $t=5.5~\text{ns}$, whereupon the signal reverts to directions
1 and 4, reaching steady state at $t=7.0~\text{ns}$. This reversible
switching process can be repeated indefinitely, as shown. We recall
that the energy-scale parameters, i.e. ${V}_{\text{HF}} |p_0|^2$,
$\gamma_x$ etc, in the polariton equations Eqs.~(\ref{Ek.equ}),
(\ref{P0.equ}), and (\ref{Pi.equ}) are of order of meV, giving a
natural dynamical time scale of ps. However, the (linear) gain and
loss rates in the off-axis modes offset each other by roughly two
orders of magnitude (the pump intensity level is set here slightly
above threshold) so that the changes in polariton density happen on
a longer time scale between several tens of ps to ns.
Another point to note is, when the control is turned on or off, it
affects not just the fields in directions (1,4) and (2,5). Every
time the control changes, the field in the third direction pair
(3,6) grows and then decays to zero. Concurrent to this movement,
the intensities in (1,4) and (2,5) undergo a slower change, forming
a shoulder on each of the time traces. These results are
qualitatively similar to those in
Ref.~\onlinecite{schumacher-etal.09}, but quantitatively
dramatically different, in that the intensity ratio of the switched
signal to the control beam in the present calculation is about two
orders of magnitude larger than in
Ref.~\onlinecite{schumacher-etal.09}. This improvement is attained
by raising the radiative decay rate of the cavity from $\Gamma = 0.4
{\rm meV}$ in Ref.~\onlinecite{schumacher-etal.09} to $\Gamma = 1.5
{\rm meV}$ here, other parameters being the same.




\begin{figure}[h]
\includegraphics[scale=0.3,angle=0,trim=00 00 00
00,clip=true]{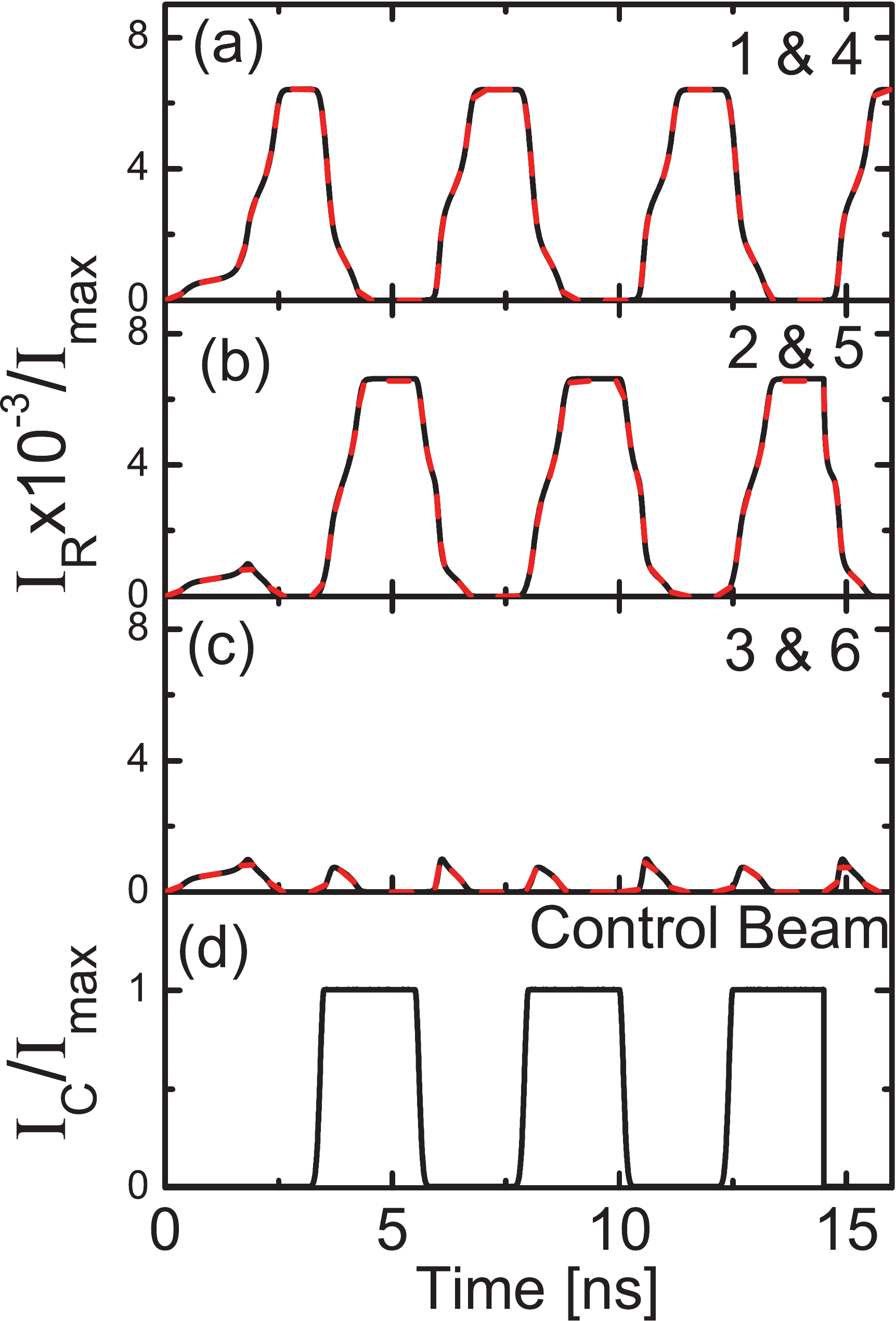} \caption{(Color online.) (a)-(c) Reflected
signals ($I_{\text{R}}$), summed over radial $\bf k$ modes, vs time
in the off-axis directions of the multi-$|{\bf k}|$ model. The
directions of the signals in each panel are marked by numerical
labels, which are defined in Fig.~\ref{sketch.fig}a. Black solid
lines represent the signals in directions 1, 2, and 3 and the red
dashed lines those in directions 4, 5, and 6. $I_{\text{max}} = 9.1 \times 10^{-3}~\text{W} \text{cm}^{-2} = 3.6 \times 10^{-7} I_{\text{pump}}$
denotes the peak control intensity. (d) The control beam intensity
($I_{\text{C}}$) vs time.} \label{erefl.fig}
\end{figure}

Fig.~\ref{timetrace.fig} shows radial ($| {\bf k} |$) distributions
of intensity in directions 1 and 2 as functions of time. At about $t
= 0.2 ~\text{ns}$, instability-generated fields appear in a range of
radial shells in all six directions. At $t \approx 1.7~\text{ns}$,
the distribution is narrowed to a single $|\mathbf{k}|$ mode, which
we label by $h_0$, in each direction.{\footnote {Under certain
conditions, instability occurs in one of the two adjacent
$|\mathbf{k}|$.}} The polariton frequency of this mode, including
the effect of the pump-induced HF and PSF shifts, is approximately
in resonance with the pump frequency. Afterwards, as shown in
Fig.~\ref{erefl.fig}, the intensity in direction 1 and $h = h_0$
rises to a steady state value while that in direction 2 falls to
almost zero. The control beam is directed at the mode with momentum
${\bf k} = {\bf k}_{h_0, 2}$. Fig.~\ref{timetrace.fig} shows the
reversible directional switching. As can be seen, there is no
spreading of the intensity distribution over the radial modes during
the switching to direction 2 or the subsequent return switching to
direction 1.


\begin{figure}[b]
\includegraphics[scale=0.3,angle=0,trim=00 00 00
00,clip=true]{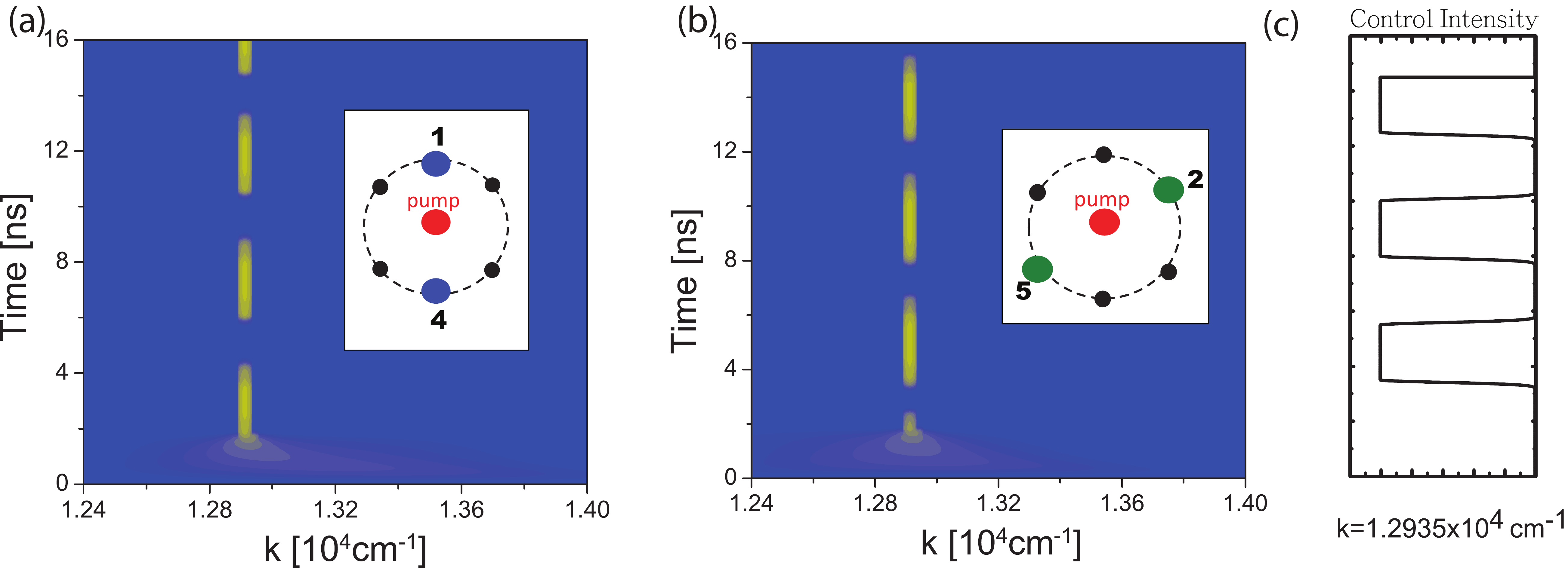} \caption{(Color online.) Radial
distributions of reflected signal intensity in momentum space in
directions 1 (a) and 2 (b) as functions of time. The intensity
distributions in directions 4 and 5 are similar to those in 1 and 2
respectively. One radial mode always `wins it all' at steady states.
The insets indicate the positions of the winning modes in a
transverse plane in the far field. (c) The temporal profile of the
control beam. In this case, the control is directed in azimuthal
direction 2 with a momentum magnitude matching that of the winning
mode in direction 1.} \label{timetrace.fig}
\end{figure}

The convergence of the off-axis signals in each direction to a
single radial momentum mode is an important feature of the
multi-$|{\bf k}|$ model simulation results. This implies that a
further reduced model, in which only one mode is included in each
hexagonal direction, would be adequate in analyzing the competitions
among patterns in a hexagonal state space and the switching process.
We call this simpler model the `single-hexagon model' and will use
it in the following sections to carry out a parameter-variation
study of the pattern competitions and analyze their underlying
physical mechanisms. In the full 2D simulations, the convergence to
one radial momentum mode is not as evident, which may be due to the
fact that the pump beam is a finite-width distribution in $\bf k$
space in the 2D simulations instead of a single mode (at ${\bf k} =
{\bf 0}$) as in the present model. Results similar to those in Fig.
~\ref{erefl.fig} have been presented in Ref.
\onlinecite{schumacher-etal.09}, but were not analyzed in detail.


\subsubsection{The ring model: results}

In the ring model, we consider modes described by
Eq.~(\ref{ring-k-pt.equ}) with $k_r = 1.2935 \times 10^4~{\rm
cm}^{-1}$, which is the winning radial momentum extracted from the
multi-$|{\bf k}|$ model (Fig.~\ref{timetrace.fig}). All parameters
are the same as in the full 2D simulations; in particular no spatial
anisotropy is given to any mode here. Low-level random seed
fluctuations in the field are put into the off-axis modes.
Instability-induced fields initially grow in all modes around the
ring, but eventually, as in the full 2D simulations, the system
stabilizes into a hexagonal pattern, the orientation of which is
arbitrary and may change from run to run. Therefore, the situation
that the symmetric ring pattern is unstable, while each `broken
symmetry' state of a hexagonal pattern is stable, is retained when
one reduces the full 2D state space to the ring configuration.

Combining the simulation results in this section, using the full 2D,
multi-$|{\bf k}|$, and ring models, one can see that an important
class of transverse instability-driven patterns exist in a hexagonal
state space. Moreover, a model including only the pump (${\bf k} =
{\bf 0}$) mode and six modes on a regular hexagon (the
single-hexagon model) would be sufficient to analyze the
competitions and controlled switching among these patterns. This
analysis will be carried out in the following sections. Interactions
among modes in different hexagons will be further studied in Section
\ref{no-quad.sec}. It will be seen that the insight gained from the
single-hexagon model is also helpful in understanding these
inter-hexagon competitions.

\section{Single-hexagon model}
\label{sinhex.sec}

In this and the following three sections, the pattern dynamics in a
hexagonal geometry are investigated in detail using the
single-hexagon model. The results for our standard parameter values
are reported in this section, and some parametric dependencies of
the results are studied in Section \ref{parameter.sec}. Detailed
analysis of the physical processes driving competition and control
is contained in Sections \ref{analysis.sec} and \ref{interplay.sec}.

\begin{figure}[b]
\includegraphics[scale=0.3,angle=0,trim=00 00 00
00,clip=true]{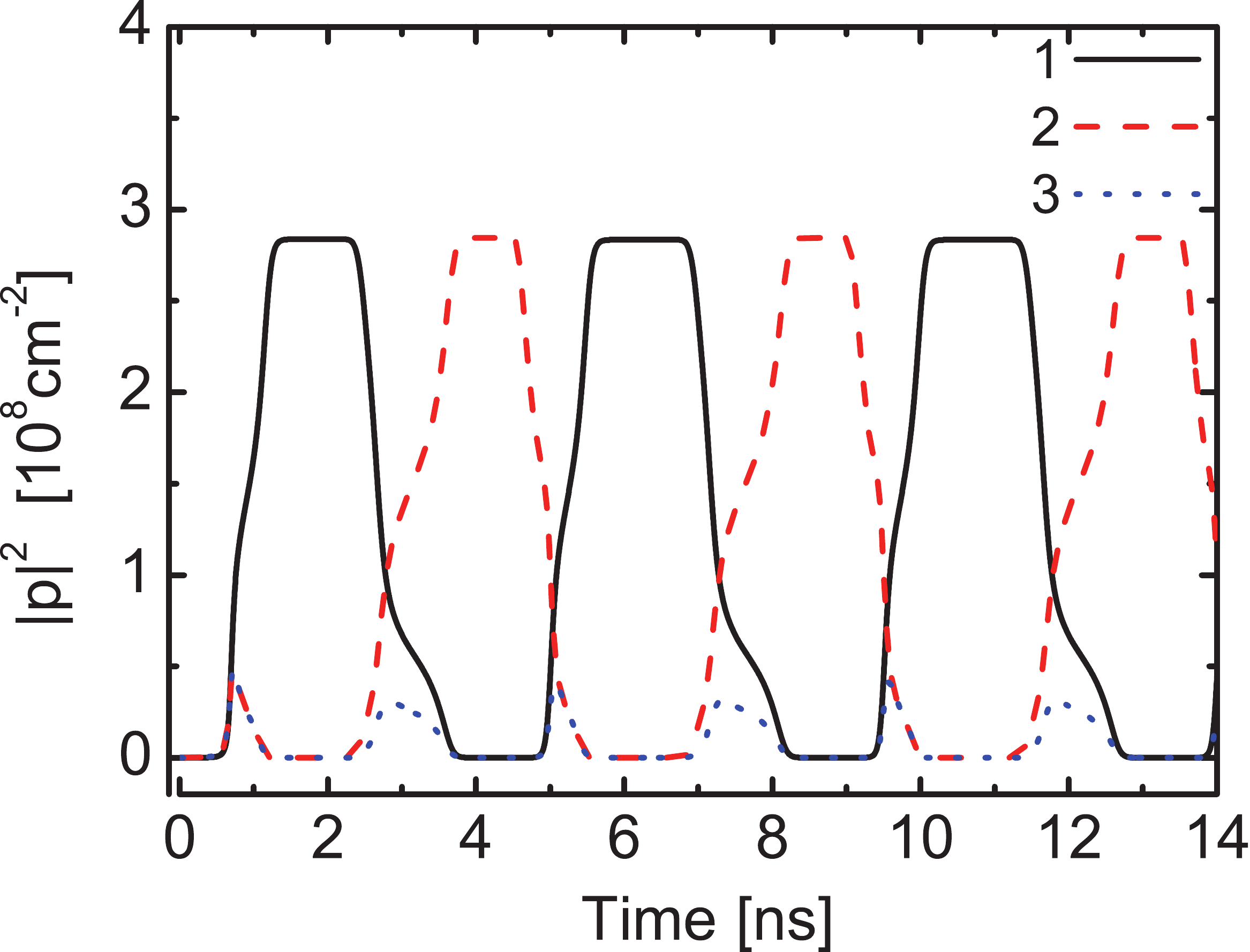} \caption{(Color online.) Exciton density
for three off-axis directions in the single-hexagon model: $|p_1|^2$
- black solid line, $|p_2|^2$ - red dashed line, $|p_3|^2$ - blue
dotted line. Note that the dotted line, $|p_3|^2$, overlaps with
the other lines over much of its range. The control beam, with peak intensity $\approx 3.6 \times 10^{-7} I_{\text{pump}}$ is first
switched on at $t = 2$ ns in this run.} \label{p1-p3.fig}
\end{figure}

\begin{figure}[b]
\includegraphics[scale=0.3,angle=0,trim=00 00 00
00,clip=true]{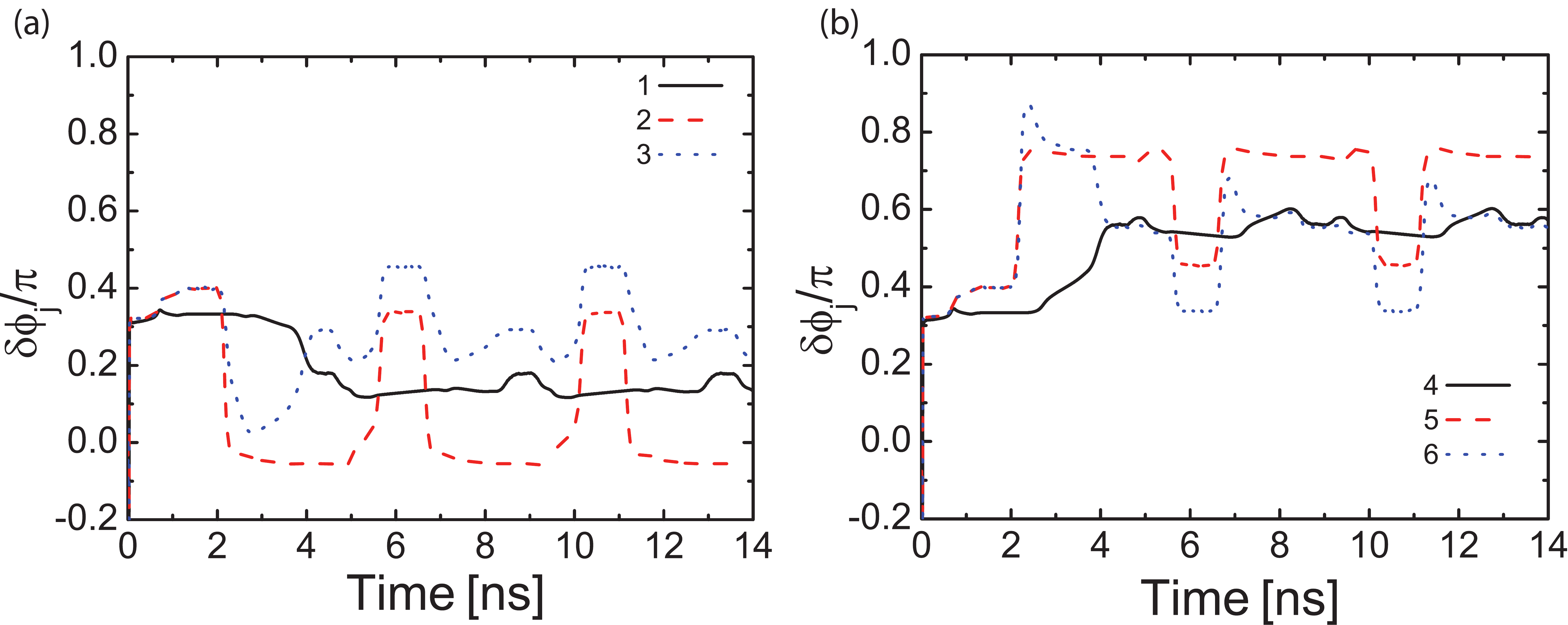} \caption{(Color online.) Phases of the
exciton fields in the off-axis directions for the reference case
within the single-hexagon model. (a) Phases in modes 1-3, and (b)
phases in modes 4-6, relative to the pump-induced exciton phase,
$\delta \phi_j = \phi_j - \phi_0, j=1-6$. The modes are labeled in
Fig.~\ref{models.fig}c.
Note that the symmetry between $\phi_j$ and $\phi_{j+3}$ is broken
after the introduction of control beam at about $t = 2$ ns.} \label{phases.fig}
\end{figure}

The single-hexagon model includes only one sextuplet of points on
the vertices of a hexagon plus the origin in $\bf k$ space, as
illustrated in Fig.~\ref{models.fig}(c). Within the labeling scheme
of the last two models we set the number of hexagons $N = 1$ and
drop the index $h$. The momenta of the off-axis modes are written as
${\mathbf{k}_{i}}=k_r \hat{\bf e}_{i} , i = 1, \cdots , 6$, where,
as in the ring model, $k_r$ is assigned the value for the winning
mode in the multi-$| {\bf k} |$ simulations. The field symbols are
also simplified to $p_i$, $E_i$, $i= 0, \cdots, 6$, with $i=0$
labeling the pump fields. Figs.~\ref{p1-p3.fig} and \ref{phases.fig}
show the simulation results using the same material and run-time
parameters as in the multi-$| {\bf k} |$ simulation presented in the
previous section. As in that simulation with the multi-$| {\bf k} |$
model, modes 1 and 4 are given a slight advantage, and a control
beam is periodically applied to mode 2. Exciton densities in three
off-axis directions, $|p_i|^2, i = 1, \cdots , 3$, are plotted as
functions of time in Fig.~\ref{p1-p3.fig}. Their behaviors are
clearly similar to those of the reflected light fields obtained
within the multi-$|{\bf k}|$ model (Fig.~\ref{erefl.fig}). In the
winning off-axis direction, the steady state exciton density is of
the order of $10^8~\text{cm}^{-2}$, or about $0.01$ of the pump
induced exciton density. While the rise time of $p_1$ in each cycle
is about the same as in the multi-$|{\bf k}|$ model results, the
time for complete switching, when the control is turned on, from
direction 1 to 2, is roughly 200 ps longer here. As explained in
the next subsection, a control intensity threshold for complete
switching exists, and the longer switching time in the
single-hexagon model indicates that this threshold is higher here
than in the multi-$|{\bf k}|$ model.


Since the polariton scatterings represented in Eqs. (\ref{P0.equ})
and (\ref{Pi.equ}) are coherent processes, the density transfer
among modes depends critically on the relative phases of the fields
involved. We display the phases of off-axis exciton fields relative
to the pump's phase in Fig.~\ref{phases.fig}. Their behaviors and
their roles in the dynamics of pattern selection will be more
thoroughly discussed in Sections \ref{analysis.sec} and
\ref{interplay.sec} below. The importance of phases was also
investigated in other instability-driven structures, e.g. in
Ref.~\onlinecite{egorov-etal.11}.

The exciton density in the pump direction ($|p_0|^2$, which is not
shown) reaches its peak value of about $10^{10}~\text{cm}^{-2}$ at
around $t = 30~\text{ps}$ and stays constant except for a slight
variation when the control is on. The phase of the pump exciton field
also stays essentially constant at roughly $0.15 \pi$ after $t =
30~\text{ps}$.

The pattern-formation time starting from the initial pump-only state varies with the models used. In general, this initial formation time is longer when there are more competing modes in the model, being shortest in the single-hexagon model and longest in the full 2D simulations. The switching time when the control beam is applied, however, is roughly the same in all models.

\section{Pattern and time scale variations with parameters}
\label{parameter.sec}

In the previous two sections, we have discussed a particular
scenario (with asymmetry applied) where a `two-spot' pattern is
selected by the system's dynamics and then the pattern is rotated by
a control beam.  In this section, we investigate the range of stable
patterns supported by the hexagonal state space and how the
selection of patterns as well as the transition time between
patterns depend on the system and control parameters. We will refer
to the set of results presented in the previous section as the
`reference case'.


\subsection{Varying the anisotropy in the cavity mode energy}

\begin{figure}[b]
\includegraphics[scale=0.3,angle=0,trim=00 00 00
00,clip=true]{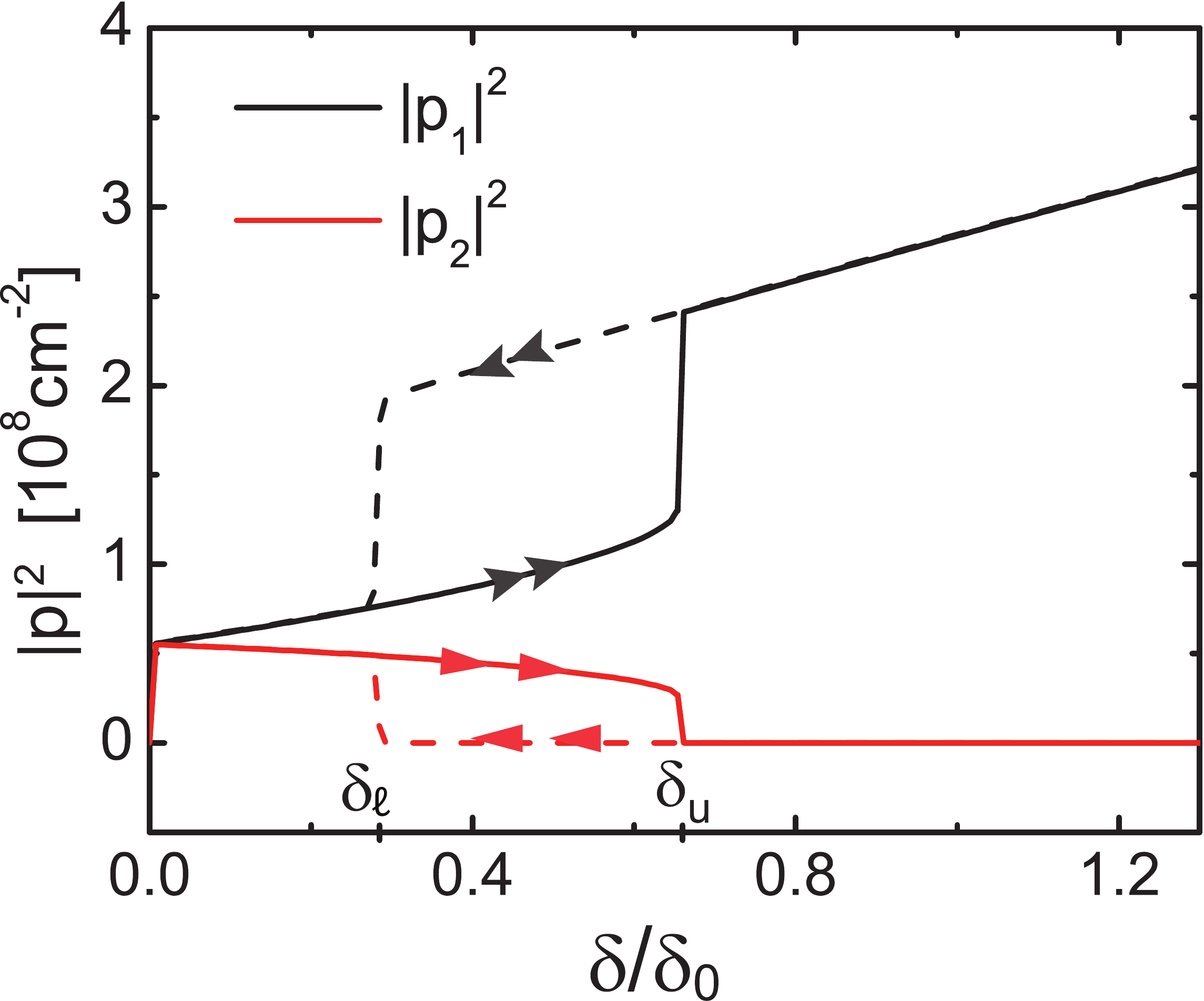} \caption{(Color online.) Steady state
exciton densities in directions 1 and 2 vs the amount of anisotropy
-- downward shift of the cavity mode energy in direction 1
($\delta$). $\delta_0 = 0.12~\text{meV}$ is the shift in the
reference case (see text). $| p_3 |^2$ (not shown) is equal to $|
p_2 |^2$. For $0 < \delta < \delta_{\ell} \approx 0.28 \delta_0$, only
the hexagonal pattern is stable. For $\delta > \delta_{u} \approx
0.66 \delta_0$, only the two-spot (in modes 1 and 4) pattern is
stable. Both patterns are stable in the range between
$\delta_{\ell}$ and $\delta_{u}$, where their adiabatic evolutions
show hysteresis behaviors: the solid (dashed) lines trace the
evolution of the densities as $\delta$ increases (decreases) from
small (large) values.} \label{pasy.fig}
\end{figure}

In the reference case, we introduced an anisotropy in the cavity
mode energy $\omega _{\mathbf{k}}^{c}$ by down-shifting it by
0.12~meV in modes (1,4). We examine here the effects of changing
the size of this anisotropy, which we denote by $\delta$.  The
reference value of 0.12~meV is denoted by $\delta_0$.
Figure~\ref{pasy.fig} shows the exciton densities in directions 1
and 2 in the steady state(s) as a function of $\delta / \delta_0$.
The density in direction 3 is equal to that in direction 2 in each
case. For $\delta = 0$, where there is complete symmetry among the
six directions in the equations, we have found only the symmetric
steady state solution, with $| p_1 | = | p_2 | = | p_3 | $, to be
stable.{\footnote {The equi-density hexagonal pattern is the
`symmetric' state in the single-hexagon model. It is, however, a
`broken-symmetry' state in the ring model where the `symmetric'
state has density evenly distributed around the ring.}} The hexagon
remains the only stable pattern for values of $\delta$ between 0 and
a lower critical value $\delta_{\ell} \approx 0.28
\delta_0$, with the steady state value of $| p_1 |$ steadily growing
while that of $| p_2 | = | p_3 |$ is falling. Between
$\delta_{\ell}$ and an upper critical value
$\delta_{u} \approx 0.66 \delta_0$, both the hexagon
and the two-spot pattern, in directions (1,4), are stable. Above
$\delta_{u}$ only the two-spot pattern is stable. In
Fig.~\ref{pasy.fig}, the solid lines trace the evolution of the
steady state densities if $\delta$ is slowly increased from zero:
they vary smoothly until $\delta$ crosses $\delta_{u}$,
when $| p_1 |$ jumps up and $| p_2 | ( = | p_3 |)$ drops abruptly to
zero. Starting from large $\delta$, when we decrease $\delta$, the
system follows the dashed lines inside the bistable range. The total
off-axis density, summed over all six directions, increases with
$\delta$ over the displayed range (not shown). For example, the
total exciton density in the symmetric case ($\delta = 0$) amounts
to about $50 \%$ of that in the reference case ($\delta =
\delta_{0}$).

\begin{figure}[b]
\includegraphics[scale=0.3,angle=0,trim=00 00 00
00,clip=true]{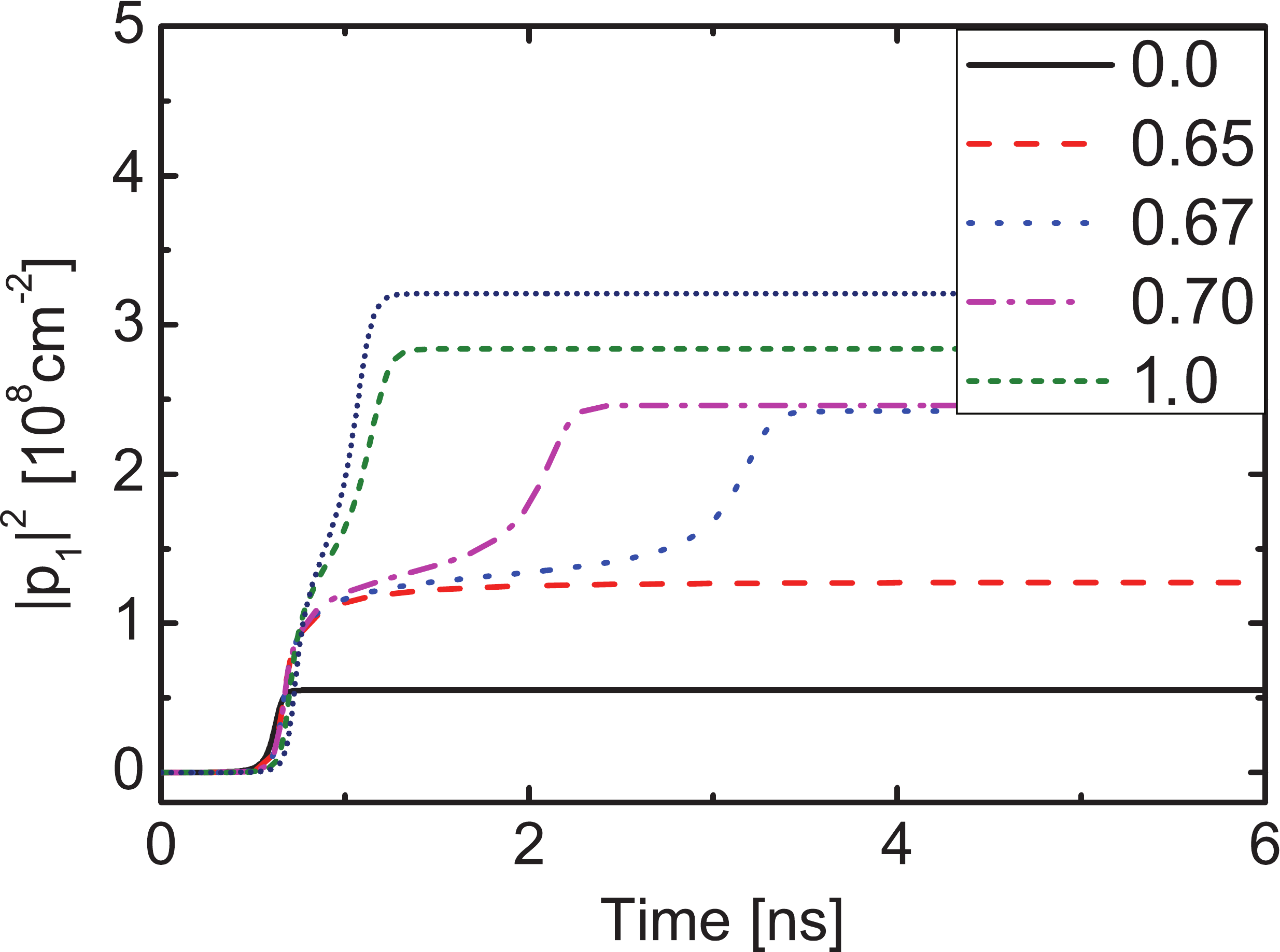} \caption{(Color online.) Exciton density
in direction 1 (preferred direction) vs time for various amounts of
downward shift in the cavity mode energy ($\delta$) calculated in
the single-hexagon model. The value of $\delta / \delta_0$ for
each curve is given in the inset. For $\delta$ below
the critical value for hexagon-to-two-spot transition (see
Fig.~\ref{pasy.fig}), $| p_1 |$ attains steady state within 1 ns.
For $\delta$ immediately above this critical value, e.g. the curve
for $\delta / \delta_0 = 0.67$, $| p_1 |$ passes through a long
intermediate step before rising to its steady state level.}
\label{p1-p3_sym.fig}
\end{figure}

In Fig.~\ref{p1-p3_sym.fig}, we compare the time evolution of
$|p_1|^2$ for various values of $\delta$ during the transition from
the initial vacuum state to the stable steady state. For this
initial state, the system goes asymptotically to the hexagon steady
state for values of $\delta$ in the bistable range. In each case,
the pump intensity becomes practically constant at around $t =
30~\text{ps}$. The off-axis density becomes visible on the scale of
the final steady state level at about 0.6 ns. After this point, for
$\delta < \delta_{u}$, $|p_1|^2$ takes a relatively
short time -- several tens of ps -- to reach its steady state value.
Beyond the hexagon-two-spot transition, however, the time trace of
$|p_1|^2$ acquires an intermediate step, during which its change
slows down, before its final rise to a constant value. The length of
this intermediate step appears to increase without limit as $\delta$
approaches $\delta_{u}$ from above. Similar to the
reference case, shown in Figs. (\ref{erefl.fig}) and
(\ref{p1-p3.fig}), for each $\delta > \delta_{u}$, $|p_2
(t)|^2$ and $|p_3 (t)|^2$ stay non-zero during the time when $|p_1
(t)|^2$ is passing through its intermediate step. Comparing the
curves at $\delta = 0.65 \delta_0$ and $\delta = 0.67 \delta_0$
suggests that the intermediate step is a `residual' effect of the
hexagonal pattern. Though no longer a stable steady state, it still
slows down the system's dynamics when the system passes through its
vicinity.
In cases not close to the hexagon-two-spot transition, one lower
bound to the formation time of the stable pattern is the ps time
scale of the electron-hole Hamiltonian. In most cases, the actual
time scale is one to two orders of magnitude longer that this lower
bound depending on the level of fluctuations and the balance between
gain and (both radiative and non-radiative) loss rates.

When the asymmetry $\delta$ is set below 0, directions 1 and 4
are given a disadvantage, and the stable pattern is a hexagon
with $|p_2|^2 = |p_3|^2 > |p_1|^2$.  A transition to a four-spot
pattern would not occur since the latter is not a steady state
solution to Eq.~(\ref{Pi.equ}) (see Section \ref{analysis.sec}
below).

The switching behavior of the hexagonal pattern for $\delta <
\delta_{u}$ is similar to that in the reference case
when the same control beam is applied. In the symmetric case
($\delta = 0$, for example), when the control beam is turned on in
direction 2, the hexagon switches to the two-spot pattern with the
density in directions 2 and 5 being the same as in the reference
case. When the control beam is then turned off, the system goes back
to the symmetric solution. The times needed for switching between
the two patterns are slightly shorter than those in the reference
case.



\subsection{Varying the control beam intensity}
In the reference case, a control beam with peak
intensity $= 9.1 \times 10^{-3}~\text{W} \text{cm}^{-2} = 3.6
\times 10^{-7} I_{\text{pump}}$ is applied in direction 2. We
examine here the effects of varying this control intensity,
retaining for all other parameters their values as in the reference
case. We denote the control intensity by $I_{\text{max}}$ and its
value in the reference case by $I^{0}_{\text{max}}$.

\begin{figure}[b]
\includegraphics[scale=0.3,angle=0,trim=00 00 00
00,clip=true]{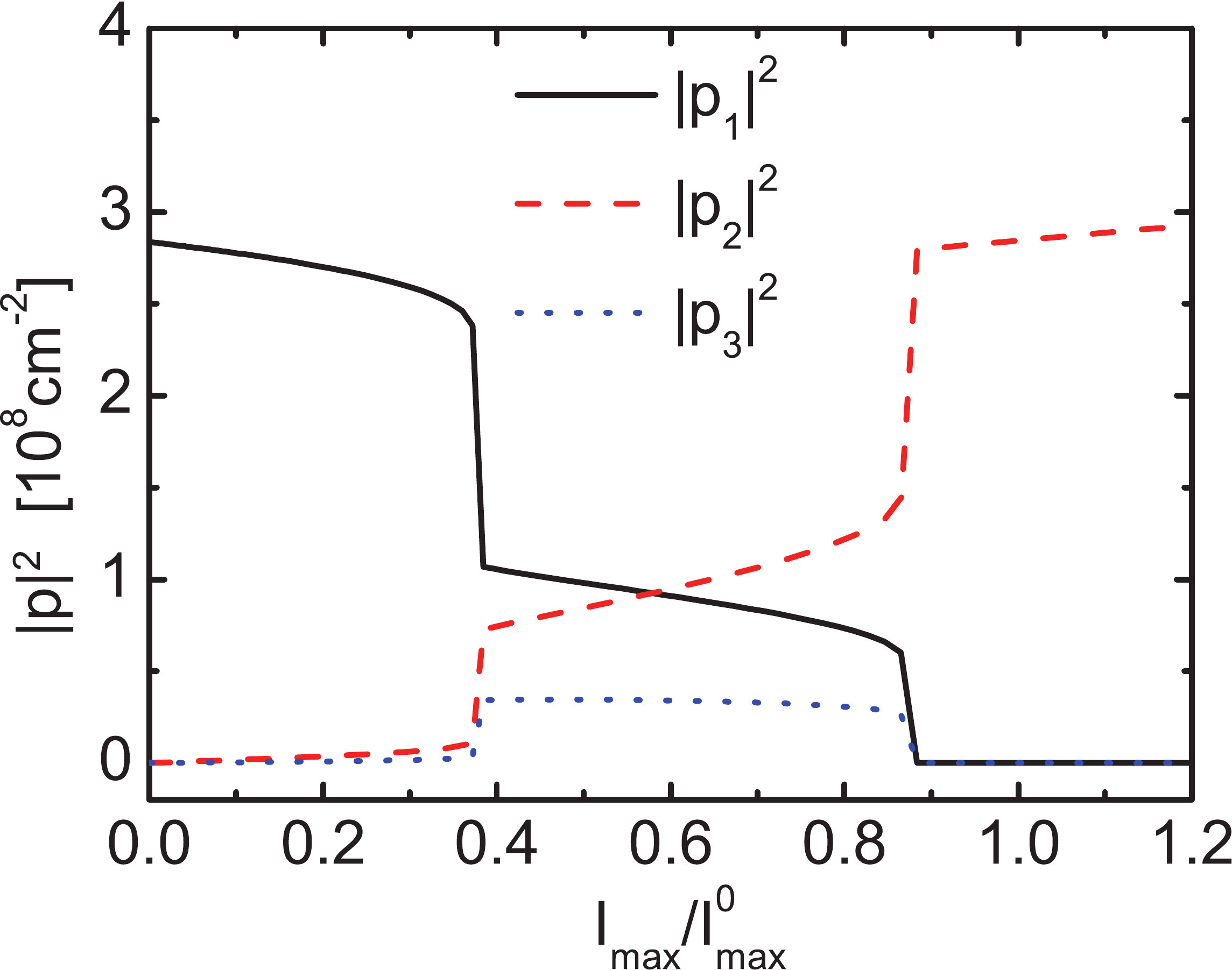} \caption{(Color online.) Steady state
exciton densities vs the peak intensity of a control beam introduced
in direction 2. The control intensity ($I_{\text{max}}$) is
expressed in units of that in the reference case, denoted here by
$I^{0}_{\text{max}}$. As $I_{\text{max}}$ increases, the pattern
goes from two spots (in directions 1 and 4) through a hexagon to two
rotated spots (in directions 2 and 5). A bistable region and
attendant hysteresis behavior (not shown in the figure) are present
at each abrupt transition. } \label{probe2.fig}
\end{figure}

Fig.~\ref{probe2.fig} shows the steady state exciton densities when
the system is under irradiation by the control beam as functions of
$I_{\text{max}}$.  It can be seen that the pattern undergoes two
abrupt changes.  When the control is off ($I_{\text{max}}=0$), the
stable pattern has two spots in directions (1,4). As
$I_{\text{max}}$ increases from zero, $|p_1|^2$ (=$|p_4|^2$) falls
while $|p_2|^2$ ($\approx |p_5|^2$) and $|p_3|^2$ (=$|p_6|^2$) rise
gradually. As $I_{\text{max}}$ reaches a critical value
$I_{\text{cr}(1)}$, which is between $0$ and $I^{0}_{\text{max}}$,
the hexagon with two bright spots and four dim spots changes
abruptly to a hexagon with comparable brightness in all spots.
Thereafter, $|p_1|^2$ continues to fall while $|p_2|^2$ continues to
grow until $I_{\text{max}}$ reaches a second critical value
$I_{\text{cr}(2)}$, which is $0.86 I^{0}_{\text{max}}$, where
$|p_1|^2$ and $|p_3|^2$ drop to zero and the pattern changes
abruptly to two spots in directions (2,5). Hence $I_{\text{cr}(2)}$
is the lowest control intensity that can completely switch the two
spots from directions (1,4) to (2,5). Though not shown in
Fig.~\ref{probe2.fig}, a bistable region and hysteresis behavior,
similar to those in Fig.~\ref{pasy.fig}, are present at each abrupt
transition.


\begin{figure}[b]
\includegraphics[scale=0.3,angle=0,trim=00 00 00
00,clip=true]{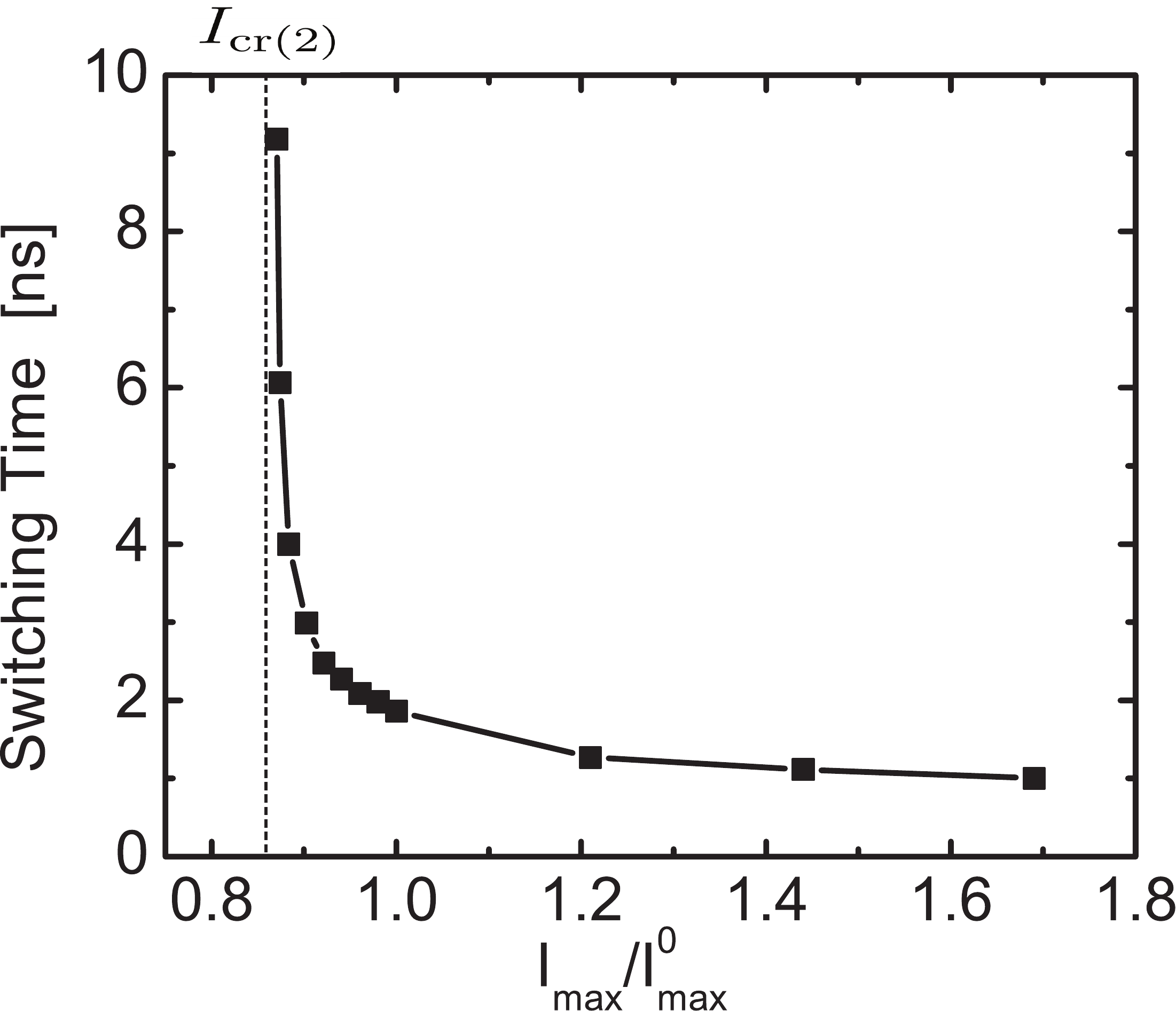}
\caption{Time taken
to switch between a two-spot pattern in directions (1, 4) to one in
directions (2, 5) vs the peak intensity of a control beam introduced
in direction 2. The control intensity ($I_{\text{max}}$) is
expressed in units of that in the reference case, denoted here by
$I^{0}_{\text{max}}$. } \label{switchtime.fig}
\end{figure}

The time traces of the densities in the reference case, in
Figs.~\ref{erefl.fig} and \ref{p1-p3.fig}, show intermediate steps
also during the switching process. As in the initial pattern
formation stage, the intermediate step here can be interpreted as a
`transient trapping' of the system in the vicinity of the hexagonal
state that loses its stability at $I_{\text{max}} =
I_{\text{cr}(2)}$. The duration of this intermediate step also
becomes longer as we set the control intensity closer to the
critical value $I_{\text{cr}(2)}$. For $I_{\text{max}}$ sufficiently
close to $I_{\text{cr}(2)}$, the time the system spends in this
transient trapping dominates the time taken by the two-spot pattern
to switch directions. Illustrating this, we plot the switching time
as a function of $I_{\text{max}}$ in Fig.~\ref{switchtime.fig}.
Again, there does not appear to be an upper bound to this duration.
Away from $I_{\text{cr}(2)}$ when the system
is no longer affected by the hexagonal state, the switching time is
around $1~\text{ns}$.

If the asymmetry $\delta$ is decreased, the critical control
intensity for rotating the two-spot pattern $I_{\text{cr}(2)}$ is
also lowered. However as there is a minimum value for $\delta$,
namely $\delta_{\text{cr}}$ for the two-point solution to be
stable, a lower bound to the switching intensity exists for a
fixed set of material parameters and pump intensity.

\section{Dynamical analysis}
\label{analysis.sec}

In the following two sections, we seek to gain insight into the
physical mechanisms underlying the phenomena of pattern formation
and switching displayed in the previous sections. For this purpose
we analyze the interplay between the various wave mixing (or
coherent scattering) processes contributing to the polariton
dynamics. To keep the discussion most transparent, we will work
within the single-hexagon model.

It is convenient to break the complex fields in the rotating frame
down to their magnitudes and phases and consider the equations of
motion for these components:
\begin{equation}
\label{ptil.equ} p_j = \tilde{p}_j e^{-i \phi_j} \, \, \, \, \,
E_j = \tilde{E}_j e^{-i (\phi_j-\xi_j)}
\end{equation}
for $j = 0, \cdot \cdot \cdot, 6$. Here $\xi_j$ is defined to be the
phase difference between the exciton field $p_j$ and the cavity
photon field $E_j$. Substituting Eq.~(\ref{ptil.equ}) into
Eqs.~(\ref{P0.equ}) and (\ref{Pi.equ}), we get for the off-axis
fields ($j = 1, \cdot \cdot \cdot, 6$),



\begin{eqnarray}
\label{p_mag.equ} \frac{d{\tilde{p}_j}^{}}{dt}&=&\frac{1}{\hbar
\tilde{p}_{j}}\operatorname{Im}[i\hbar {\dot{p}_j}p_j^{*}] =
L_{p_j} + Q_{p_j}+C_{p_j}
\\
\label{phase.equ} \frac{d\phi_j}{dt} &=& \frac{1}{\hbar
{\tilde{p}_j}^2} {\rm Re}[i\hbar {\dot{p}_j}p_j^{*}] = \frac {1}
{\tilde{p}_j} \left[ L_{\phi_j} + Q_{\phi_j}+C_{\phi_j} \right]
\end{eqnarray}
where
\begin{eqnarray}
\label{L-mag.equ} \hbar L_{p_j} &=& V_{\text{HF}}\tilde{p}_{j+3}
\tilde{p}_{0}^{2} \sin{(\Phi_{j,j+3}^{0,0})}  + 2 \tilde{A} \Omega
\tilde{p}_{0} \tilde{E}_{0} \tilde{p}_{j+3}
\sin{(\Phi_{j,j+3,\xi_0}^{0,0})} \nonumber \\&& - \Omega
\tilde{E}_{j} \sin{(\xi_j)} ( 1 - 2 \tilde{A} \tilde{p}_{0}^2) + 2
\tilde{A} \Omega \tilde{p}_0 \tilde{E}_0 \tilde{p}_j \sin{(\xi_0)}-
\gamma_x \tilde{p}_j
\\
\label{Q-mag.equ}\hbar Q_{p_j} &=&
2\tilde{p}_{j-2}\left[V_{\text{HF}}\tilde{p}_0
\tilde{p}_{j-1}\sin{(\Phi_{j, j-2}^{0, j-1})}+\tilde{A} \Omega
\tilde{E}_0\tilde{p}_{j-1}\sin{(\Phi_{j,j-2,\xi_0}^{0,j-1})} \right.
\nonumber \\&& \left. + \tilde{A} \Omega
\tilde{p}_0\tilde{E}_{j-1}\sin{(\Phi_{j,j-2,\xi_{j-1}}^{0, j-1})}
\right] + 2\tilde{p}_{j+2} \left[
V_{\text{HF}}\tilde{p}_0\tilde{p}_{j+1}\sin{(\Phi_{j,j+2}^{0, j+1})}
\right. \nonumber \\ &&+ \left. \tilde{A} \Omega \tilde{E}_0
\tilde{p}_{j+1}\sin{(\Phi_{j,j+2,\xi_0}^{0, j+1})}+
\tilde{A}\Omega\tilde{p}_0\tilde{E}_{j+1}\sin{(\Phi_{j,j+2,\xi_{j+1}}^{0,
j+1})} \right] \nonumber \\&& + 2\tilde{p}_0 \left[
V_{\text{HF}}\tilde{p}_{j+1}\tilde{p}_{j-1}
\sin{(\Phi_{j,0}^{j+1,j-1})}+\tilde{A} \Omega
\tilde{E}_{j+1}\tilde{p}_{j-1}\sin{(\Phi_{j,0,\xi_{j+1}}^{j+1,j-1})}
\right. \nonumber
\\&& + \left. \tilde{A} \Omega \tilde{p}_{j+1}
\tilde{E}_{j-1}\sin{(\Phi_{j,0,\xi_{j-1}}^{j+1,j-1})} \right]
\\
\label{C-mag.equ} \hbar C_{p_j} &=& 2V_{\text{HF}} \tilde{p}_{j+3}
\left[ \tilde{p}_{j+1}\tilde{p}_{j-2}\sin{(\Phi_{j,j+3}^{j+1,j-2})}+
\tilde{p}_{j+2}\tilde{p}_{j-1}\sin{(\Phi_{j,j+3}^{j+2,j-1})} \right]
\nonumber\\ &&+ 2 \tilde{A} \Omega  \tilde{p}_{j+3}
\sum\limits_{i\ne j,j+3} \tilde{p}_i \tilde{E}_{i+3}
\sin{(\Phi_{j,j+3,\xi_{i+3}}^{i,i+3})}\nonumber\\&& + 2 \tilde{A}
\Omega \tilde{E}_j \sin{(\xi_j)} \sum\limits_{i} \tilde{p}_i^2 + 2
\tilde{A} \Omega \tilde{p}_j \sum\limits_{i\ne j} \tilde{p}_i
\tilde{E}_{i} \sin{(\xi_i)}
\end{eqnarray}
and
\begin{eqnarray}
\label{phase-L.equ} \hbar L_{\phi_j} &=& \left[
\varepsilon_{h,j}^{x}-\hbar \omega_p + V_{\text{HF}}
\tilde{p}_{0}^{2} \right] \tilde{p}_{j} +
V_{\text{HF}}\tilde{p}_{j+3} \tilde{p}_{0}^{2}
\cos{(\Phi_{j, j+3}^{0, 0})} \nonumber
\\&& + 2 \tilde{A} \Omega \tilde{p}_{0} \tilde{E}_{0}
\tilde{p}_{j+3} \cos{(\Phi_{j,j+3,\xi_0}^{0, 0})} \nonumber \\ && -
\Omega \tilde{E}_{j} \cos{(\xi_j)} \left[ 1 - 2 \tilde{A}
\tilde{p}_{0}^2 \right] + 2 \tilde{A} \Omega \tilde{p}_0 \tilde{E}_0
\cos{(\xi_0)} \tilde{p}_{j} \\
\hbar Q_{\phi_j} &=& \hbar Q_{p_j} [\sin \mapsto \cos ] \label{phase-Q.equ} \\
\hbar C_{\phi_j} &=& \hbar C_{p_j} [\sin \mapsto \cos ] +
V_{\text{HF}} \tilde{p}_j (2\sum\limits_{i \neq j} \tilde{p}_i^2+
\tilde{p}_j^2) \label{phase-C.equ}
\end{eqnarray}
where $\Phi_{i,j}^{l,m} =
\phi_i+\phi_j-\phi_l-\phi_m$ and $\Phi_{i,j,\xi_k}^{l,m} =
\phi_i+\phi_j+\xi_k-\phi_l-\phi_m$. In these equations, each sum
over $i$ ranges from 1 to 6. In Eqs.~(\ref{phase-Q.equ}) and
(\ref{phase-C.equ}), a function followed by the symbol $[\sin
\mapsto \cos ]$ means the same function with all sines replaced by
cosines in its expression. We have grouped the contributing terms
according to their order in the off-axis fields. As the symbols
imply, $L_{p_j}$, $Q_{p_j}$, and $C_{p_j}$ contain respectively
first (linear), second (quadratic), and third-order (cubic) terms in
$p_j$ or $E_j$, $j = 1, \cdot \cdot \cdot, 6$. The same applies to
$L_{\phi_j}$, $Q_{\phi_j}$, and $C_{\phi_j}$. The state with
polariton density only in the pump mode is treated as the
zeroth-order state. The pump fields $p_0$ and $E_0$ experience very
small (fractional) changes in time and will be taken approximately
as constant in this analysis.





As mentioned before, all exciton interactions and PSF terms can be
visualized as either a four-wave mixing process with wave-vector
matching or a polariton-polariton scattering process with momentum
conservation. For each scattering process in Eqs.~(\ref{P0.equ}) and
(\ref{Pi.equ}), the exciton field on the left-hand-side and the
conjugated field on the right-hand-side are in the outgoing modes
while the other two fields on the right-hand-side are in the
incoming modes. Of course these scatterings are coherent, not of the
Boltzmann, mass-action type: the direction of density transfer
depends on the interference between the scattered wave and the
existing field in each outgoing channel. In
Eqs.~(\ref{p_mag.equ})-(\ref{phase-C.equ}), the phases of the
incoming fields are subtracted from those of the outgoing fields
inside the sine or cosine functions. Since the density equations
contain the sine functions, the term is a gain term for the mode in
question if the combined phases of the outgoing fields lead (by less
than $\pi$) those of the incoming fields. Otherwise, it is a loss
term. For the phase's rate of change, with the cosine functions, a
process tends to increase the phase in question if the phase
difference between the outgoing fields and the incoming fields is
less than $\pi / 2$.

\begin{figure}[b]
\includegraphics[scale=0.3,angle=0,trim=00 00 00
00,clip=true]{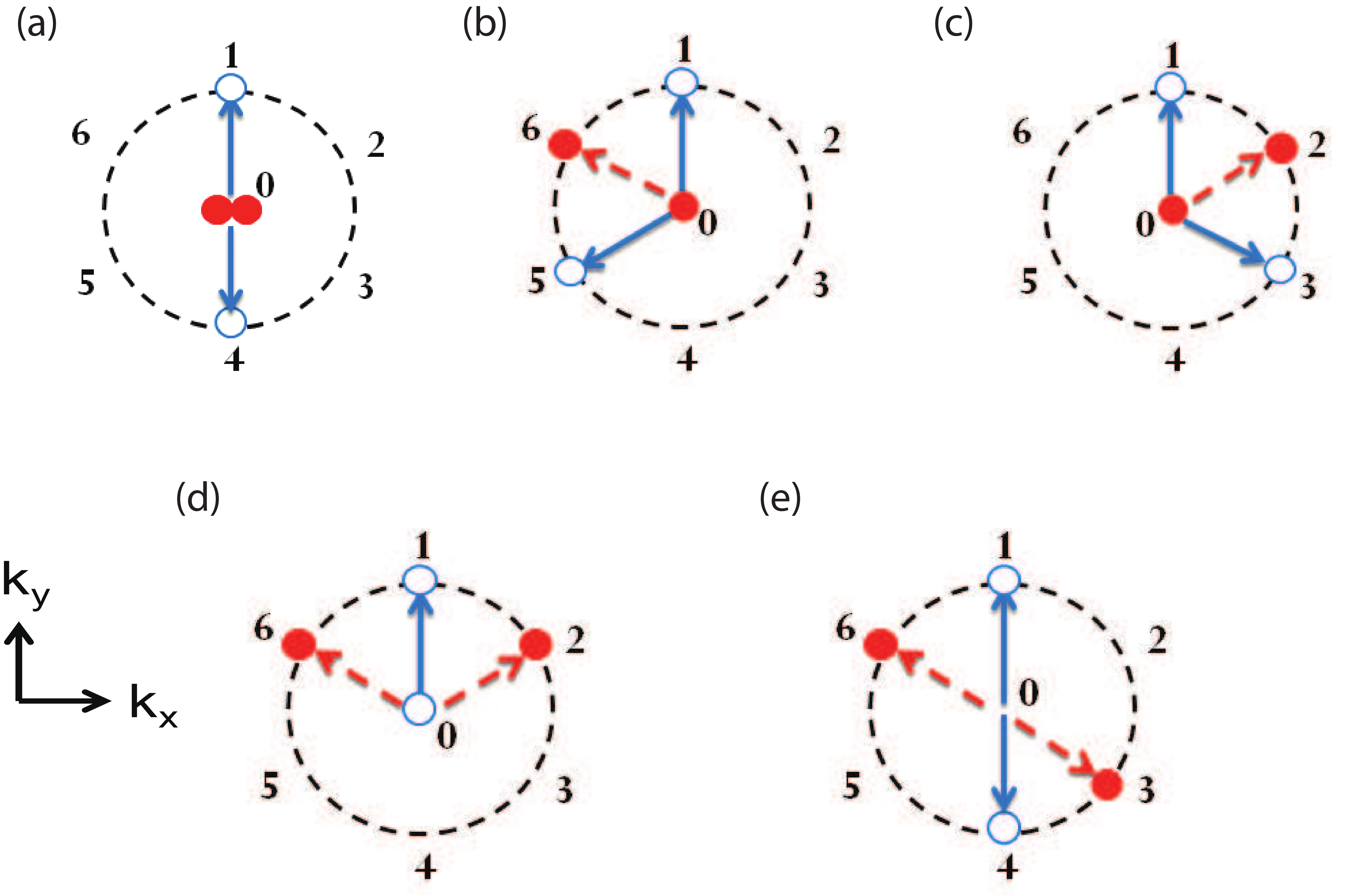} \caption{(Color online.) Pictorial
representations of polariton scattering processes contributing to
the equation of motion of $p_1$. The processes are (a) linear, (b),
(c), (d) quadratic, and (e) cubic in the off-axis fields. In each
diagram, the incoming modes and outgoing modes in the scattering are
labeled by red, solid circles and blue, open circles respectively.
The arrows represent the modes' momenta, dashed (solid) for the
incoming (outgoing) modes. For example in (b), two polaritons from
directions 0 (pump) and 6 scatter into directions 1 and 5.}
\label{picofterms.fig}
\end{figure}

The first two terms contributing to $L_{p_j}$ in
Eq.~(\ref{L-mag.equ}) are the primary instability-driving (gain)
processes, in which two pump polaritons scatter off each other into
two opposite off-axis directions. An example is displayed
pictorially in Fig.~\ref{picofterms.fig}(a). The next term is the
exciton-photon coupling in an off-axis mode, partially blocked by
the pump exciton density. In the calculations, the angle $\xi_j$ is
always positive, corresponding to an expected net energy flow from
the exciton field $\tilde{p}_j$ to the photon field $\tilde{E}_j$.
Hence, to the exciton field, this term represents the radiative loss
from mode ${\bf k}_j$. The next term describes the forward
scattering between a pump photon and an off-axis exciton, or
equivalently, a self-wave-mixing process in which a pump photon
scatters off a density grating, set up by the pump and the off-axis
exciton fields, into the off-axis direction. In the pump mode,
energy flows on balance from the cavity photon field to the exciton
field, resulting in a negative phase-lag angle $\xi_0$. Hence, this
scattering actually results in a density transfer from the off-axis
${\bf k}_j$ mode to the pump mode. The last term $- \gamma_x
\tilde{p}_j$ represents all non-radiative losses. The balance
between gain and loss terms determines the sign and magnitude of
$L_{p_j}$.

In a quadratic process, i.e. one that contributes to $Q_{p_j}$ and
$Q_{\phi_j}$, one off-axis polariton (from mode ${\bf k}_j$) and a
pump polariton scatter into the modes ${\bf k}_{j-1}$ and ${\bf
k}_{j+1}$ and vice versa. These processes are also illustrated in
Fig.~\ref{picofterms.fig}: the first three terms in
Eq.~(\ref{Q-mag.equ}) are represented by
Fig.~\ref{picofterms.fig}(b), the next three terms by
Fig.~\ref{picofterms.fig}(c), and the last three terms by
Fig.~\ref{picofterms.fig}(d). We note that the momentum conservation
condition that gives rise to the quadratic terms is satisfied only
for modes arranged geometrically in a hexagon. Among the cubic
terms, the first three contributing to $C_{p_j}$ and $C_{\phi_j}$
represent two counter-directed polaritons on the hexagon scatter
into a different pair of opposite directions. An example is drawn in
Fig.~\ref{picofterms.fig}(e). The next two terms are, respectively,
the Pauli blocking reduction of the (linear) exciton-photon coupling
by the exciton density in all off-axis modes and a self-wave-mixing
process in which a photon in mode $i$ ($i \neq j$) scatters off a
density grating, set up by $p_i$ and $p_j$, into mode $j$. Since $0
< \xi_i < \pi/2$ for $i = 1, \cdot \cdot \cdot , 6$, these two terms
are positive in both Eqs.~(\ref{C-mag.equ}) and (\ref{phase-C.equ}).
The last term contributing to $C_{\phi_j}$ in
Eq.~(\ref{phase-C.equ}) is a (blue) shift in frequency for $p_j$. In
contrast to the quadratic terms discussed above, these cubic
processes do not require a hexagon mode configuration to be
operative.


We finally note a phase degeneracy in the solution to
Eqs.~(\ref{p_mag.equ})-(\ref{phase-C.equ}). Suppose $({\tilde p}_j
(t), \phi_j (t)), j = 1, \cdots, 6$ is a solution to these
equations. Then one can verify that $({\tilde p}_j (t), \phi_j (t)+
\delta \phi_j)$ is also a solution provided the six $\delta
\phi_j$'s are time-independent and satisfy the following
constraints:
\begin{eqnarray}
\label{phaseconserve1.equ} \delta \phi_j + \delta \phi_{j+3} &=&
2\pi m
\\
\label{phaseconserve2.equ} \delta \phi_{j+1} + \delta \phi_{j-1} -
\delta \phi_{j}&=& 2\pi m
\end{eqnarray}
where $m$ is an integer, and the subscript is again counted
cyclically through 1 to 6. These two
constraints restrict the number of undetermined phases to two. This
applies in particular to steady state solutions. Thus there exist an
infinite number of steady state solutions to
Eqs.~(\ref{p_mag.equ})-(\ref{phase-C.equ}) which can be grouped into
a finite number of classes. The solutions in each class have the
same set of magnitudes and are parameterized by two free phases. We
see an example of this phase freedom in Fig.~\ref{phases.fig}, when
we compare the phases of, say, $p_1$ and $p_4$ during the initial
steady state with their phases when the system returns to the (1,4)
two-dot pattern after the first on/off switching cycle. The
individual steady state phases are different in the two time
periods, while their sum stays the same. Therefore, strictly
speaking, the system `returns' after the on/off switching cycle to a
different steady state that has the same set of magnitudes as before.
Put another way, in a hypothetical experiment, the measured relative
phase between $E_1$ and $E_0$ may be affected by uncontrollable
factors, but the measured average phase of $E_1$ and $E_4$ relative
to $E_0$'s phase can be meaningfully compared to our theory's
prediction.


%

\section{Interplay of wave mixing processes}
\label{interplay.sec}

We now discuss the contributions and interplay of the various
processes in effecting pattern selection and switching in the
reference case (Section \ref{sinhex.sec}). It is clear from
Eqs.~(\ref{L-mag.equ})-(\ref{C-mag.equ}) that to understand how the
exciton densities evolve, one needs information on the relative
phases of the polariton fields involved in the scatterings. The
calculated magnitudes and phases of the off-axis $p_j$ for the
reference case were plotted in Figs.~\ref{p1-p3.fig} and
\ref{phases.fig}. Figure~\ref{allterm.fig} shows the contributions
of the three groups of terms -- linear, quadratic, and cubic -- to
the rates of change of the exciton densities in directions 1 and 2,
and Fig.~\ref{allphterm.fig} shows these contributions to the
evolution of the respective exciton phases.
Initially, the linear terms dominate, and in each off-axis mode, the
exciton phase $\phi_j$ is quickly locked into a value given by the
solution of ${d\phi_j} / {dt} \approx {L_{\phi_j}} /
{\tilde{p}_j}=0$, where ${L_{\phi_j}}$ is given by
Eq.~(\ref{phase-L.equ}). Actually, as explained above, this
condition fixes only the value of the sum $\phi_j + \phi_{j+3}$ but
not either phase angle individually. In the calculations, however,
all the exciton phases are initially set at 0, and since
${L_{\phi_j}} / {\tilde{p}_j} \approx {L_{\phi_{j+3}}} /
{\tilde{p}_{j+3}}$, we also have $\phi_j \approx \phi_{j+3}$ in the
linear regime. The radial $| {\bf k} |$ value of the off-axis modes
having been chosen optimal, we have (cf Eq. (\ref{L-mag.equ}))
\begin{equation}\label{phase-lin-rel.equ}
\Phi_{j, j+3}^{0, 0} = \phi_j + \phi_{j+3} - 2 \phi_0 \approx \pi /
2.
\end{equation}
In Fig.~\ref{phases.fig}, this phase lead is about $0.6 \pi$ for
each of the three off-axis mode pairs. Once their phases are locked,
the off-axis exciton densities grow exponentially from fluctuation
levels. We are considering pump intensities close to the
phase-conjugate instability threshold. This condition implies a
substantial cancelation between the gain and loss terms in
$L_{p_j}$. In the reference case calculations, the exponential
off-axis growth rate in the first 0.6 ns is about 0.01 ${\rm
ps}^{-1}$, which is much slower than the typical inverse time scales
given by the exciton Hamiltonian (e.g. $\gamma_x / \hbar = 0.67~{\rm
ps}^{-1}$). During this exponential growth phase, the ${\tilde
p}_j$'s are actually too small to be visible on the scale in
Fig.~\ref{p1-p3.fig}. It is interesting to note that, although the
asymmetry-favored $p_1$ (and $p_4$) eventually wins out at steady
state, $p_2$ actually grows slightly faster than $p_1$ during the
initial 0.6 ns. Thus the asymmetry confers its advantage through the
quadratic and/or cubic processes. Put another way, the selection of
($p_1$, $p_4$) as the winning directions by the asymmetry works
through the interactions among excitons in all six off-axis
directions.

\begin{figure}[b]
\includegraphics[scale=0.3,angle=0,trim=00 00 00
00,clip=true]{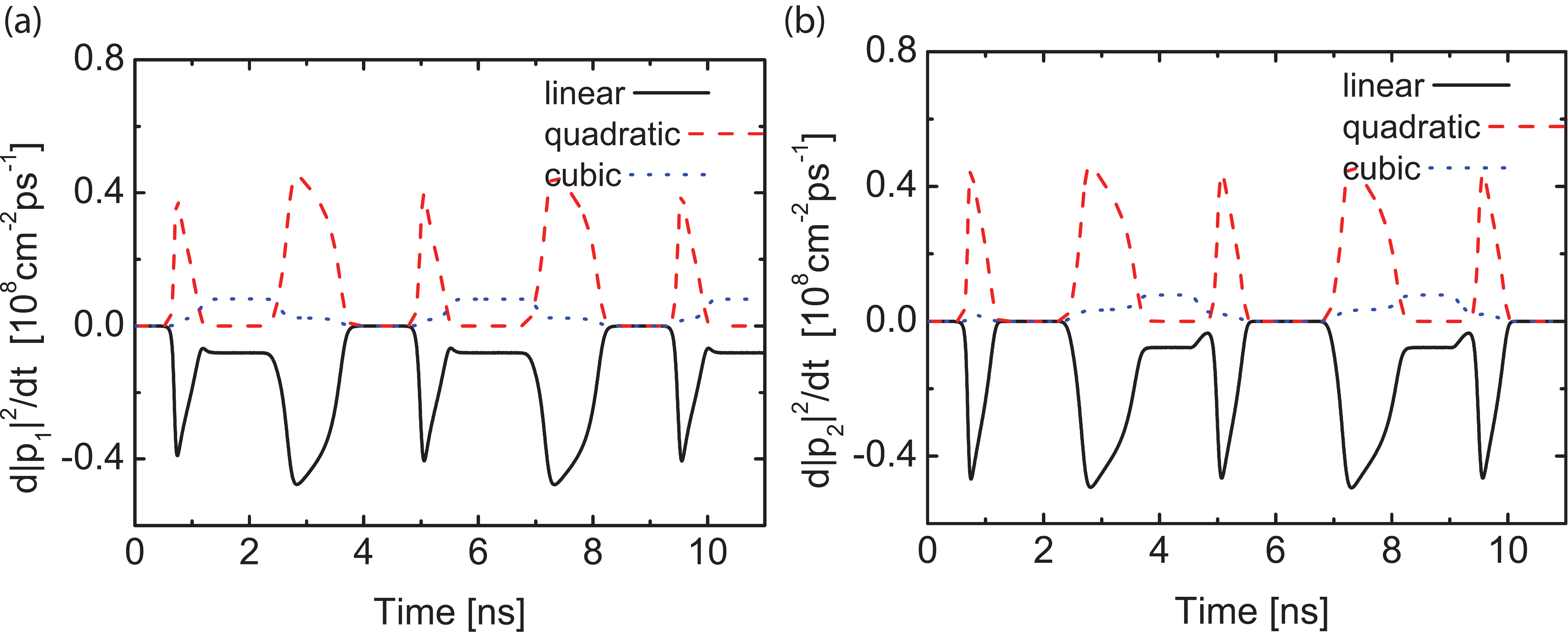} \caption{(Color online.) Contributions of
the linear, quadratic, and cubic terms to the rate of change of
exciton densities in directions 1 (a) and 2 (b) in the reference
case.} \label{allterm.fig}
\end{figure}

\begin{figure}[b]
\includegraphics[scale=0.3,angle=0,trim=00 00 00
00,clip=true]{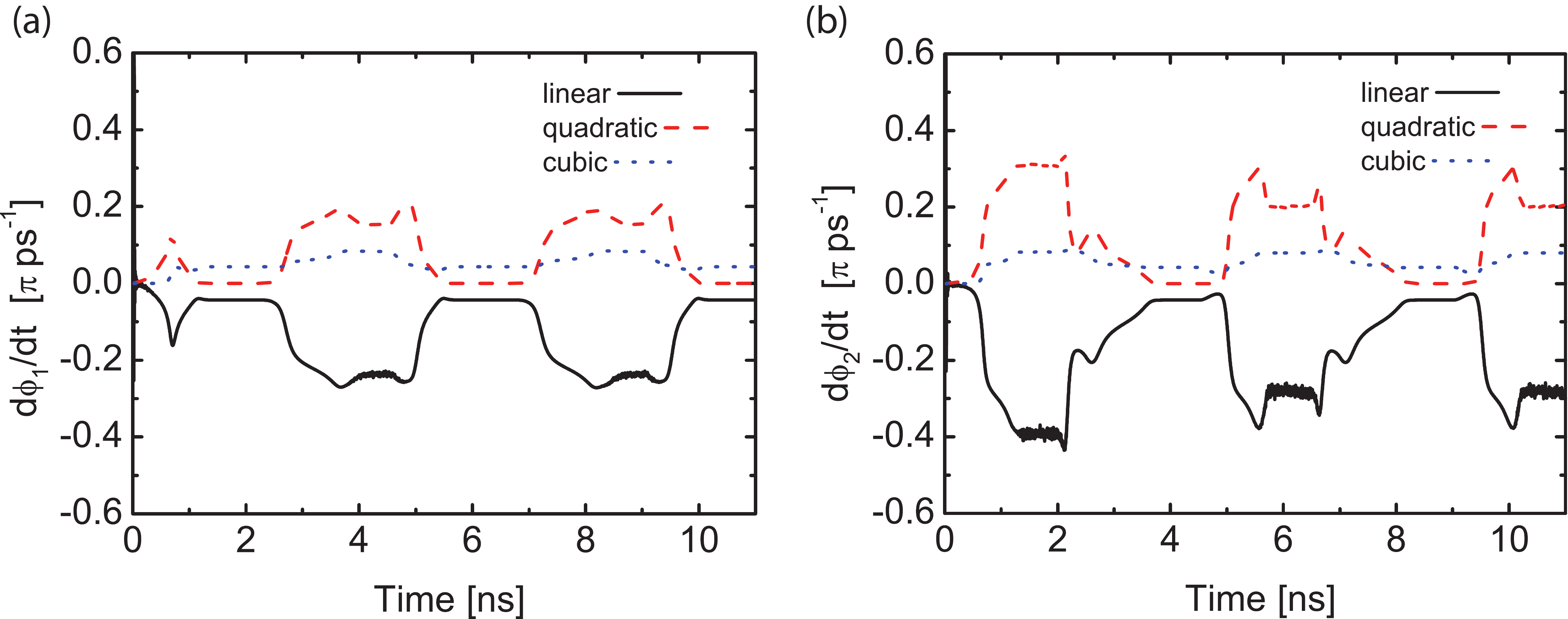} \caption{(Color online.) Contributions
of the linear, quadratic, and cubic terms to the rate of change of
exciton phases in directions 1 (a) and 2 (b) in the reference case.}
\label{allphterm.fig}
\end{figure}

The linear growth regime lasts approximately 0.5 ns. As the
$\tilde{p}_j$'s grow, the quadratic and cubic contributions to $ d
p_j / d t$ come into play. The onset of the quadratic terms has two
effects. First, their {\it direct} contribution to the magnitude
equation, $Q_{p_j}$ in Eqs.~(\ref{p_mag.equ}) and (\ref{Q-mag.equ}),
is positive, as shown in Fig.~\ref{allterm.fig}. This can be
understood as follows. As explained in Section \ref{analysis.sec},
whether a scattering term is a gain or loss process depends on the
phase sum of the outgoing exciton-polariton fields relative to that
of the incoming fields. In each term contributing to $Q_{p_j}$, one
of the four exciton fields is in the pump mode while the other three
are off-axis. As we have seen above in
Eq.~(\ref{phase-lin-rel.equ}), the linear instability sets up the
off-axis fields with phases leading the pump's by a large margin. In
the first six terms of Eq.~(\ref{Q-mag.equ}), represented by
Fig.~\ref{picofterms.fig}(b) and (c), $p_0$ is among the incoming
fields, and thus the relative phase angles are typically positive
(between 0 and $\pi$). These six terms are therefore gain terms. The
photon/exciton phase shift in each mode, $\xi_j$, is very small and
its effect can be ignored in this argument. In contrast, the last
three terms in Eq.~(\ref{Q-mag.equ}), represented by
Fig.~\ref{picofterms.fig}(d), yield losses because $p_0$ is outgoing
in them. Since the three triplets of terms are roughly equal in
magnitude, and the gains outnumber the losses, the net result for
$Q_{p_j}$ is typically positive.

The second effect of the quadratic terms acts through their
contributions to ${d\phi_j} / {dt}$ in Eq.~(\ref{phase.equ}). Again
from the linear phase lead, Eq.~(\ref{phase-lin-rel.equ}), one can
deduce that the relative phases in the terms contributing to
$Q_{\phi_j}$ in Eq.~(\ref{phase-Q.equ}) typically lie, during the
onset of the quadratic terms, between $- \pi /2$ and $\pi / 2$,
$Q_{\phi_j}$ is therefore positive, as shown in
Fig.~\ref{allphterm.fig}. The resulting increase in $\phi_j$ exerts
a negative feedback effect on the linear scattering process in
$L_{p_j}$: since the relative phase in the linear gain terms ${\rm
sin} ( \Phi_{j, j+3}^{0, 0} )$ is a little over $\pi / 2$ during the
initial growth period, pushing $\phi_j$ (and $\phi_j{+3}$) up
reduces the gain. And since the loss terms, e.g. $\gamma_x
\tilde{p}_j$, are not affected, $L_{p_j}$ suffers a net decrease and
becomes negative, as can be seen in Fig.~\ref{allterm.fig}. In fact,
the upsurge of ${\tilde p}_{1/4}$ to their winning (two-spot) steady
state is associated with a slower rise in $\phi_{1/4}$ than in the
other modes' phases. Initially locked at zero, $L_{\phi_j}$ is also
decreased by the growing $\Phi_{j, j+3}^{0, 0}$ negative values, as
shown in Fig.~\ref{allphterm.fig}, thus offsetting the effects of
$Q_{\phi_j}$ to some degree.

The contributions of the third-order scattering processes to the
exciton density, i.e. the first three terms in
Eq.~(\ref{C-mag.equ}), depend on the relative phases between the
off-axis fields. When ${\tilde p}_{1}$ is winning, since $\phi_1 <
\phi_2 = \phi_3$, inspecting Eq.~(\ref{C-mag.equ}) shows that these
terms are negative for $C_{p_1}$ and positive for $C_{p_2}$ and
$C_{p_3}$, or they tend to equalize the densities in the six modes.
The remaining two terms in $C_{p_j}$ are Pauli blocking terms that
reduce the flow of density from $p_j$ to $E_j$ and so are gain terms
to mode $j$. In the two-spot steady state with modes 1 and 4, the
only surviving contribution to $C_{p_1}$ is $2 \tilde{A} \Omega
\tilde{E}_1 \tilde{p}_1^2 \sin{(\xi_1)}$. Similar to $Q_{\phi_j}$,
all terms in $C_{\phi_j}$ in Eq.~(\ref{phase-C.equ}) are positive,
thus again helping to lower the linear amplitude term $L_{p_1}$. The
first set of terms in Eq.~(\ref{phase-C.equ}) again represent
coherent scatterings and Pauli blocking. The last term is a blue
shift of the exciton energy in mode $j$ due to interactions with
off-axis excitons.


The above analysis underlines the fact that pattern competition in
even the single-hexagon model is the result of rather complex,
activating or inhibiting, feedbacks among the various physical
processes. Not being simply additive, the effects of the processes
are not easy to disentangle cleanly. Nevertheless, it is instructive
to consider an artificial reduction of the single-hexagon model in
which the quadratic processes are disabled. In this case, the most
basic function of the cubic processes is to saturate the polariton
density in the off-axis modes, which the cubic processes achieve
mostly through raising the phases of $p_j$'s, as explained above.
Setting the quadratic terms (last three lines in the single-hexagon
of Eq. (\ref{Pi.equ}), or equivalently, $Q_{p_j}$ and $Q_{\phi_j}$
in Eqs. (\ref{p_mag.equ}) and (\ref{phase.equ})) to zero, we have
repeated the simulations with the same parameters as in the runs
shown in Section \ref{parameter.sec}. Fig.~\ref{p1-p3_ns.fig} shows
the exciton densities in modes 1 to 3 for an isotropic setting, i.e.
no advantage is given to modes 1 and 4 ($\delta =0$). Unlike the
`physical' case, with the quadratic terms operative, where the
symmetric hexagon pattern is stable, the stable steady state in
Fig.~\ref{p1-p3_ns.fig} is a two-spot pattern -- the six-fold
symmetry of the single-hexagon model is spontaneously broken. Which
pair of modes win at the expense of the others is not determined by
the model's physics and changes from run to run. When a sufficiently
strong control beam is applied to mode 2, the pattern switches over
to modes 2 and 5, but stays in this pair when the control is turned
off. This is consistent with the fact that, in the isotropic
setting, the two-spot pattern at modes 2 and 5 is the stable state
closest to the steady state maintained by the control beam.

\begin{figure}[b]
\includegraphics[scale=0.3,angle=0,trim=00 00 00
00,clip=true]{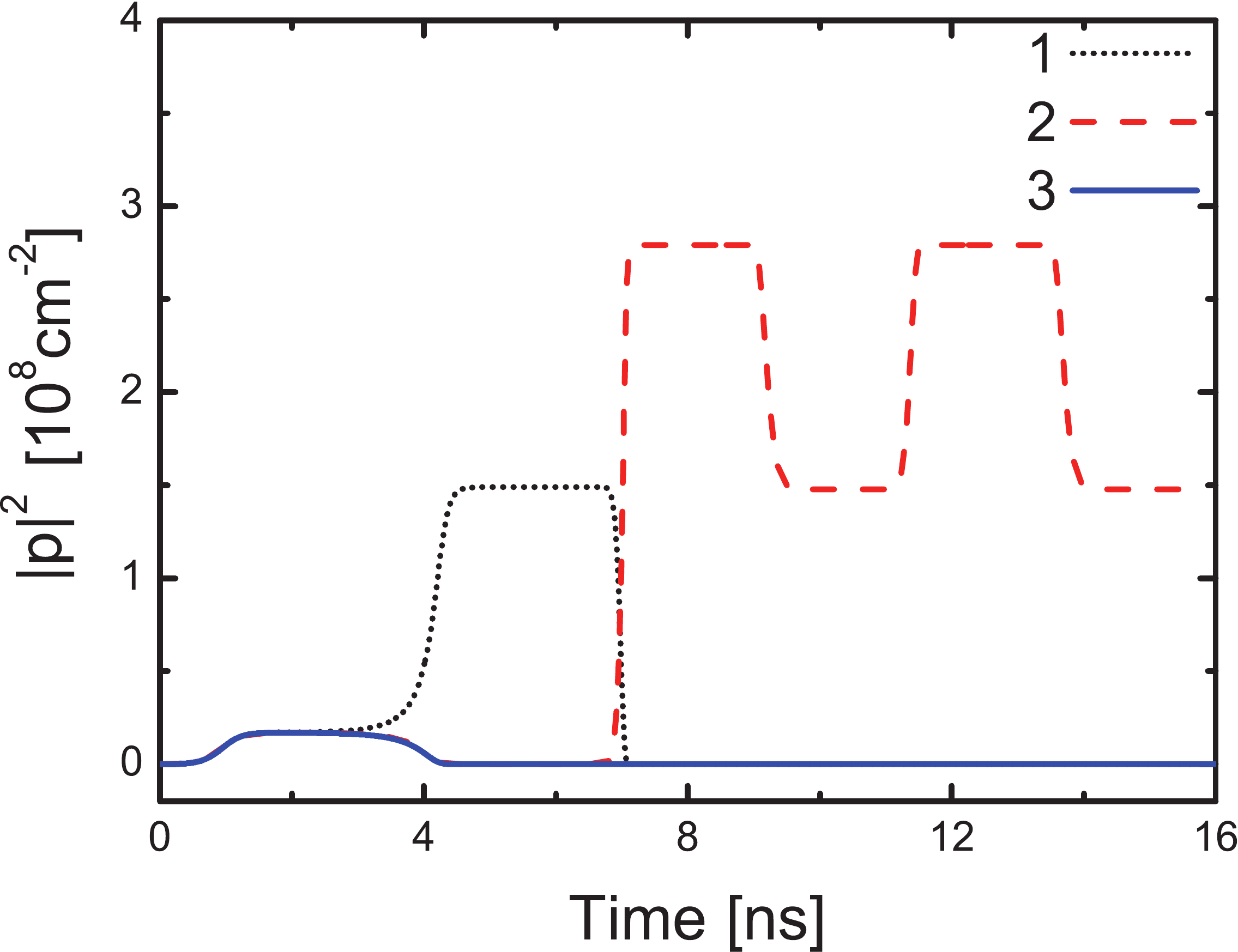} \caption{(Color online.) Exciton
densities in directions 1 to 3 for the reference case but with the
quadratic terms disabled and no asymmetry in the cavity
dispersion relation. The first on/off cycle of the control beam
starts at $t = 7 \text{ns}$.}
\label{p1-p3_ns.fig}
\end{figure}

It is useful here to draw some insight from previous analyses of
other multiple-mode competition models\cite{lamb.64,ciliberto-etal.90,longhi-geraci.98} that are based on linear
growth and cubic saturation. Consider Lamb's two-mode laser
model, \cite{lamb.64} where each mode exerts a cubic saturation
effect on both itself and the other mode. For this model, a
symmetric steady state, with equal intensity in the two modes, is
stable if self-saturation is more efficient than cross-saturation,
while two asymmetric states, each with intensity in only one mode,
are stable if cross-saturation is more efficient. By analogy, we
also subdivide the contributions to $C_{\phi_j}$ and $C_{p_j}$ into
those due to densities in modes in $j$ and $j+3$ and those due to
densities in other modes, and refer to the two classes as
`self-saturating' and `cross-saturating' contributions respectively.
The analogy is not exact because the effects of the two classes on
$d {\tilde p}_j / dt$ are not simply additive, as their counterparts
in Lamb's model \cite{lamb.64} are, but these arguments are still
applicable qualitatively, and our numerical results indicate that
the cross-saturation cubic processes in our model polariton system
are more efficient than the self-saturation processes. We have
already seen that turning the quadratic processes back on in the
isotropic setting destabilizes the two-spot pattern while restoring
stability to the symmetric hexagonal pattern. This indicates that
the enhancement brought by $Q_{p_j}$ to $d {\tilde p}_j / dt$
outweighs the suppression caused by the positive $Q_{\phi_j}$ in
raising the phase of $p_j$. Thus the net effect of the quadratic
processes can be classified as `cross-activating'. The introduction
of anisotropy ($\delta > 0$) modifies the balance of the above
activating and saturating effects and, when strong enough,
destabilizes the hexagonal pattern.

In the above, we have analyzed the physical processes driving
transverse pattern selection and control in a quantum well
microcavity and attempted to pursue a semi-phenomenological analogy
between the pattern dynamics in this system and other mode-(or
pattern-)competition systems. We will explore this analogy further
in a future publication.

\section{Competitions among hexagons}\label{no-quad.sec}


\begin{figure}[b]
\includegraphics[scale=0.3,angle=0,trim=00 00 00
00,clip=true]{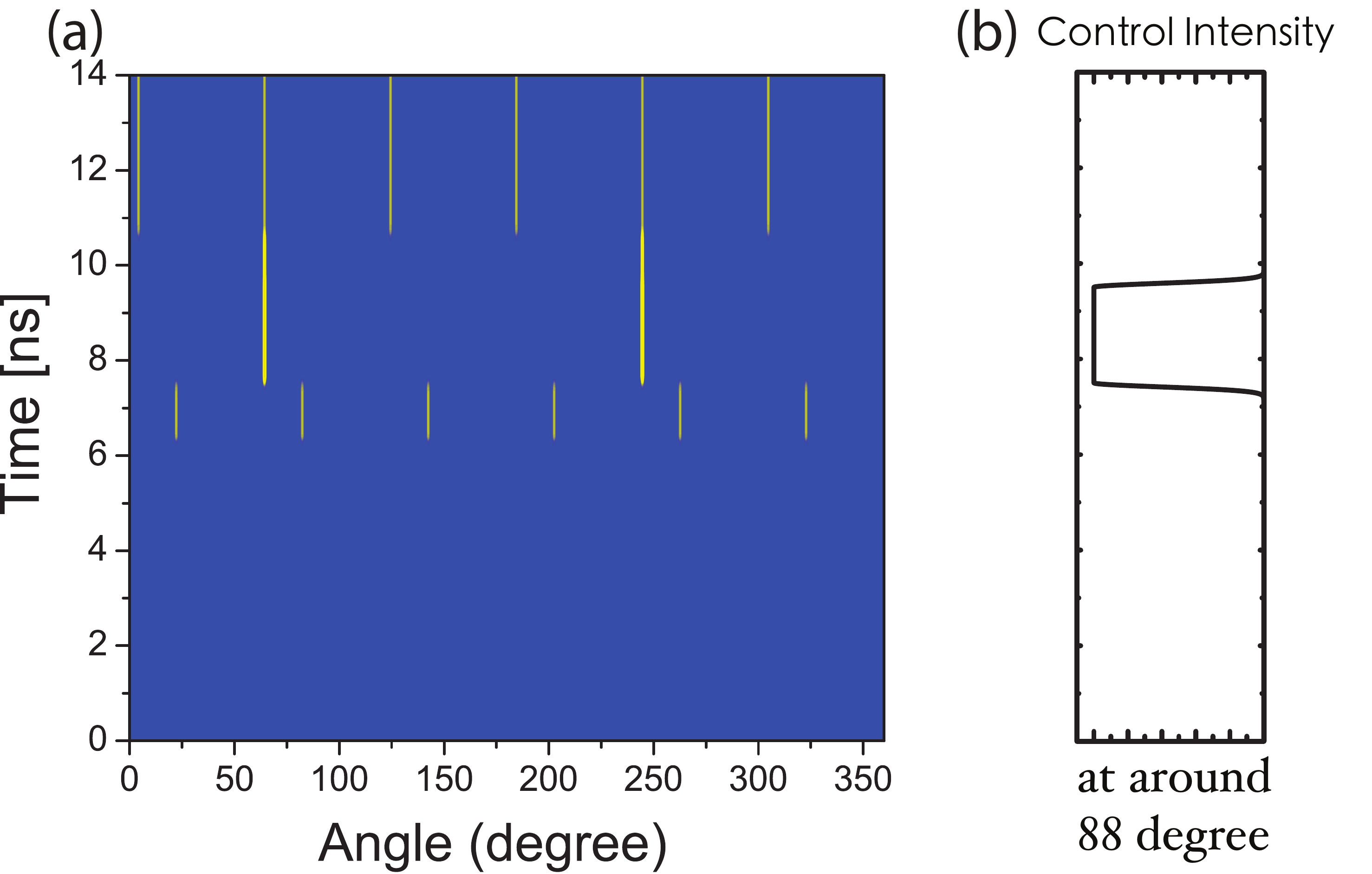} \caption{(Color online.) (a) Time
evolution of the angular distribution of the reflected signal
intensity in momentum space, calculated in the ring model. Although
the setup is perfectly isotropic, the ring pattern is unstable while
the hexagons are stable. The orientation of the hexagonal pattern
obtained in each run is arbitrary. A control beam is introduced at
$t= 7 \text{ns}$, directed at a mode outside of the initial hexagon,
switching the pattern to a two-spot one. When the control beam is
switched off at $t=9.5 \text{ns}$, the signals form a hexagon in the
orientation defined by the control beam. (b) Temporal profile of the
control beam intensity.} \label{ring.fig}
\end{figure}

\begin{figure}[b]
\includegraphics[scale=0.3,angle=0,trim=00 00 00
00,clip=true]{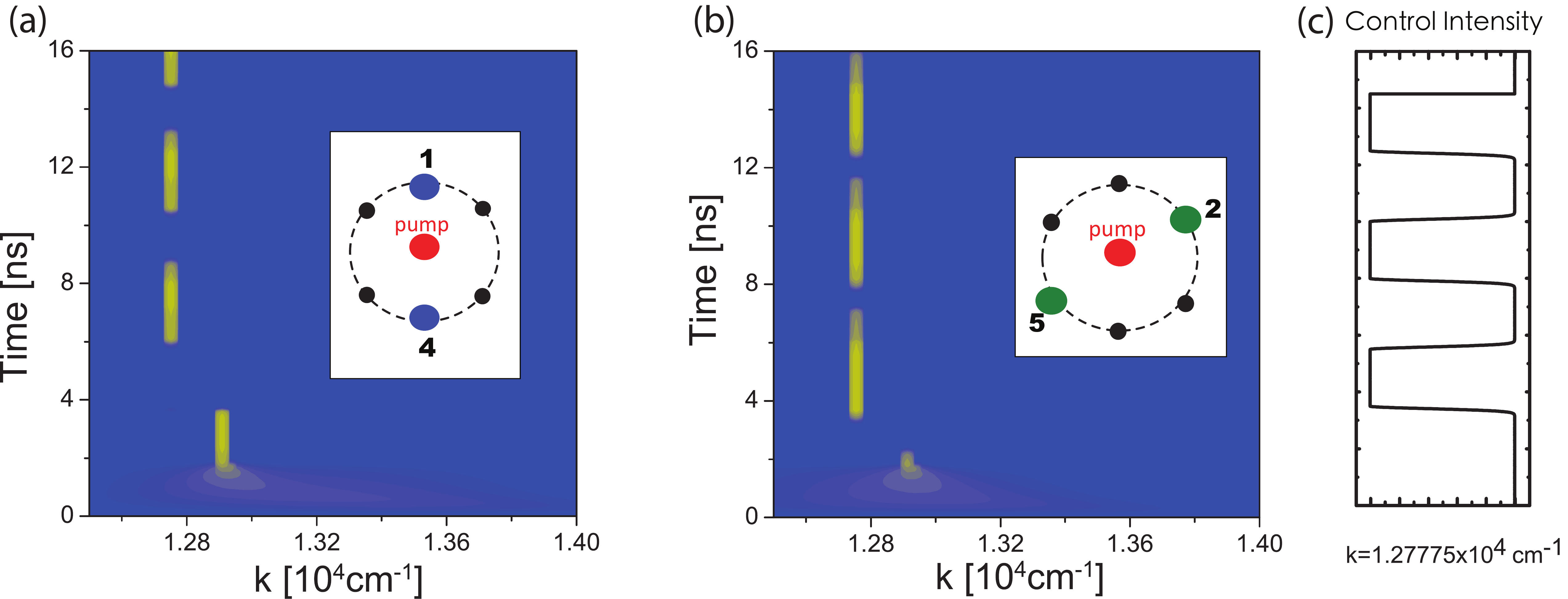} \caption{(Color online.) Time
evolution of the radial distributions of reflected signal intensity
in momentum space in directions 1 (a) and 2 (b), calculated in the
multi-$|{\bf k}|$ model. One radial mode always `wins it all' at
steady state. The inset in each panel indicates the positions of the
winning modes in a transverse plane in the far field. (c) Temporal
profile of the control beam. In this case, the control is directed
in azimuthal direction 2 with a momentum magnitude smaller than that
of the winning mode in direction 1.} \label{timetrace_shifted.fig}
\end{figure}

With the single-hexagon model, we have analyzed the switching
between patterns that are subsets of the same (regular) hexagon. In
this section, we investigate switching between patterns that reside
in different hexagons, as displayed in simulations using the
multi-$|\bf k|$ and ring models. One can see by inspecting Eqs.
(\ref{Pi.equ}) that the quadratic processes (last three lines of
Eqs. (\ref{Pi.equ})) take place only among modes situated on the
same hexagon. Modes on different hexagons compete with each other
only through the cubic processes. From the intuition that we formed
in the previous section, that cubic cross-saturation in our system
is more efficient than self-saturation, we expect patterns that
break the symmetry \emph{among} hexagons to be favored. We have
already seen that this is consistent with the simulation results in
the 2D $\bf r$-space model and the ring model in Section
\ref{models.sec}. In both simulations, the setup is isotropic, but a
hexagon pattern with arbitrary orientation is stable while the ring
pattern is not. Within the hexagon, the cross-activating quadratic
processes stabilize the symmetric hexagonal pattern at the expense
of the two-spot solutions. In Fig.~\ref{ring.fig} we show some
extended results of the ring model simulation. The figure shows the
time evolution of the angular distribution of intensity.  Starting
with only the pump mode, the system forms a hexagon at about $t
\approx 6.5 \text{ns}$, and the exciton density in the winning
hexagon is $5.6 \times 10^7~{\rm cm}^{-2}$, which is the same as the
steady state density in the symmetric multi-$|{\bf k}|$ model (not
shown). A control beam with the same intensity as the one in the
reference case is applied to another mode on the ring. Just as in
the reference case, the control stabilizes a two-spot pattern with
one spot in the mode it shines on. When the control beam is
subsequently turned off, the off-axis pattern reverts to a hexagon
which is oriented to contain the previous two-spot pattern as a
subset. This is again consistent with the fact that each individual
hexagon pattern is stable, and when the two-spot pattern is rendered
unstable by the control beam's removal, the system settles into the
nearest stable pattern.

A comparable situation is present in the multi-$|{\bf k}|$
simulations. We performed a simulation similar to the one shown in
Figs. \ref{erefl.fig} and \ref{timetrace.fig} except that the
control beam is applied to a mode on a different hexagon (i.e.,
having a different radial momentum $|{\bf k}|$) from the initial
two-spot pattern. Recall the notations in Section \ref{models.sec}:
with the anisotropy-induced preference, the initial two-spot pattern
is in modes ${\bf k}_{h_0,1}$ and ${\bf k}_{h_0,4}$, and the control
beam is applied to mode ${\bf k}_{h_0,2}$. Here the control is
directed at mode ${\bf k}_{h'_0,2}$  with $h'_0$ being a close
neighbor of $h_0$. Fig.~\ref{timetrace_shifted.fig} shows the time
evolution of the exciton density in modes over a range of $h$, or
$|{\bf k}|$, values in directions 1 and 2. One can see that when the
control is on, the pattern follows it to the mode pair ${\bf
k}_{h'_0, 2}$ and ${\bf k}_{h'_0, 5}$. When the control is turned
off, the pattern switches back to directions 1 and 4 but stays in
the radial momentum mode set by the control, i.e. to the modes ${\bf
k}_{h'_0,i} ,i = 1,4$ instead of the original ${\bf k}_{h_0,i} ,i =
1,4$. In subsequent cycles of switching, the pattern stays in the
hexagon defined by $h'_0$. These behaviors again indicate that, in
the absence of the control, there exist a group of stable steady
states, in each one of which the signal is concentrated in a mode
with ${\bf k} = {\bf k}_{h,i} ,i = 1,4$ and $h$ is within a certain
range about $h_0$. Which
state the system evolves into depends on the system's history, e.g.
which mode the signal is in when the control is turned off. In
contrast to the situation in the ring model, however, there is no
symmetry under which these modes are equivalent: for example, they
have different single-polariton energies.

\section{Summary and Outlook}
\label{summary.sec}

In this paper, we have studied the nonlinear polariton dynamics of a
laser-pumped quantum well microcavity as a spontaneous
pattern-forming system. Using a microscopic theory of
exciton-exciton interactions and GaAs parameters, we have examined
the formation, selection, and control of the various transverse
polariton density modulation patterns. We have analyzed in detail
the polariton scattering processes that govern the competitions
among patterns, obtaining useful insights into the crucial role
played by the (relative) phases of the polariton amplitudes in
driving the gains and losses of the densities in various $\bf k$
modes. We have also performed a more comprehensive investigation of
a previously proposed low-intensity, all-optical switching
scheme\cite{schumacher-etal.09} that exploits these patterns.


One outcome of our analysis is a qualitative characterization of the
effects of the various scattering processes in terms of notions used
in discussing pattern formation and competition in other, usually
classical, physical systems. These notions are based only on the
polariton densities, despite the fact that the competitions among
polariton amplitudes acts to an important extent through their
phases. This has helped us form an intuitive picture of the pattern
dynamics in our polariton system. We have pursued this analogy with
other pattern-forming systems further and have constructed a
three-state mode-competition model which can be related to our
models here and whose dynamics mimic almost all the qualitative
behaviors displayed so far by the solutions of the single-hexagon
model. Because of its simplicity, this mode-competition model is
amenable to semi-analytic treatment, and we have obtained a complete
classification of the steady state solutions in the relevant region
of parameter space. This `global' perspective introduces a useful
organizational framework to discuss the qualitative dynamics of our
system. This work will be reported in a future publication.

\section{Acknowledgements}
The Paderborn group acknowledges financial support from the DFG and
a grant for computing time at $\mathrm{PC^2}$ Paderborn Center for
Parallel Computing. The CUHK group thanks the assistance and the
computing facilities available from the Research Computing Team at
The Chinese University of Hong Kong. N. H. K. acknowledges financial
support from the Center of Optical Sciences of the Chinese
University of Hong Kong. We also thank J. Lega, T. Y. Tsang, and K.
P. Chan for very helpful discussions.


\appendix

\section{Coupling between the cavity modes and macroscopic fields}
\label{field-coupling.sec}

In this appendix, we give more details on our model for the coupling
between the macroscopic light field and the cavity modes. The
relations Eqs.~(\ref{Eeff.equ}) - (\ref{Erefl.equ}) between input
and output fields will be derived, and the choice of Eq.
(\ref{Ek.equ}) will be given support by showing it leads to the
interpretation of the cavity field magnitude squared $| E_{\bf k}
|^2$ as a photon density. In our model, the cavity modes are taken
to be the spatial Fourier modes of an oscillator field separate from
the macroscopic light field and confined to an infinitely thin plane.
The QW is assumed to be embedded inside this planar
cavity, and the oscillator field and the exciton field are coupled
to form the polariton field. The (linear) coupling between the
cavity oscillator field and the outside radiation field is modeled
like the coupling between the latter and a material oscillator
field.
Since all macroscopic light modes involved are near-normal (deviations
of their wave vectors from the normal direction are typically less
than 10 degrees), for simplicity, the complications due to the
finite polar angles of the off-axis modes are ignored.
More realistic
modeling of the microcavity and its coupling to macroscopic light
field would lead to quantitative corrections to our results but are
not expected to change any conclusions in this paper.

The propagation of the macroscopic light field is governed by the
Maxwell equations (with no external `free' charges or currents). The
medium outside the cavity is taken as a dielectric with a real
refractive index $n_s = \sqrt{\epsilon_s}$ and magnetic permeability
$\mu_s = 1$. The oscillator field acts as a local contribution to
the polarization density, yielding the following constitutive
equations:
\begin{eqnarray}
{\bf {\cal D}} ({\bf r} , t) & = & \epsilon_0 n^2_s {\bf {\cal E}}
({\bf r} , t)
- \hbar t_c {\bf E}_{\rm cav} (x,y,t) \delta(z) \label{D-P.equ}\\
{\bf {\cal H}} ({\bf r} , t) & = & \frac {1} {\mu_0} {\bf {\cal B}}
({\bf r} , t) \label{H-B.equ}
\end{eqnarray}
Here the functions ${\bf {\cal E}}$, ${\bf {\cal D}}$, ${\bf {\cal
B}}$, and ${\bf {\cal H}}$, denote the positive-frequency parts of
the respective fields so that, e.g., the (real-valued) electric
field is given by $2 {\rm Re} {\bf {\cal E}}$. ${\bf E}_{\rm cav}
(x,y,t)$ is the amplitude of the resonance mode inside the cavity
whose equation of motion is given below (and in Ref.
\onlinecite{schumacher-etal.09}), and $t_c$ is a parameter
characterizing the coupling between the light field and the
oscillator mode. Substituting Eqs.~(\ref{D-P.equ}) and
(\ref{H-B.equ}) into the Maxwell equations and eliminating ${\bf
{\cal B}}$ in the usual way, we obtain a wave equation for the
radiation electric field ${\bf {\cal E}}$:
\begin{equation}
\frac {1} {\mu_0} \nabla \left( \nabla \cdot {\bf {\cal E}} \right)
+ n^2_s \epsilon_0 \frac {\partial^2}{\partial t^2} {\bf {\cal E}} -
\frac {1} {\mu_0} \nabla^2 {\bf {\cal E}} = \hbar t_c \delta (z)
\frac {\partial^2}{\partial t^2} {\bf E}_{\rm cav}
\end{equation}
For near-normal fields, $\nabla \cdot {\bf {\cal E}} \approx 0$
since $\nabla \cdot {\bf {\cal E}}$ is exactly zero at any point
outside the cavity, and, across the cavity, it is proportional to
the difference in $E_z$ which is small.  We take the approximation
$\nabla \cdot {\bf {\cal E}} = 0$ in the above equation, and
replacing $\epsilon_0 \mu_0$ by $1/c^2$ we obtain
\begin{equation} \label{E-wave.equ}
\frac {n^2_s} {c^2} \frac {\partial^2}{\partial t^2} {\bf {\cal E}}
- \nabla^2 {\bf {\cal E}} = \frac {\hbar t_c} {\epsilon_0 c^2}
\delta (z) \frac {\partial^2}{\partial t^2} {\bf E}_{\rm cav}
\end{equation}
We expand the fields in their
spatial Fourier modes in the cavity's plane, i.e., in the $(x,y)$
directions:
\begin{eqnarray}
{\bf {\cal E}} ({\bf r} , t) & = & \frac {1} {{\cal L}^2} \sum_{\bf
k} e^{i (k_x x + k_y y)} {\bf {\cal E}}_{\bf k} (z,t)
\label{transverse-Fourier.equ} \\ {\bf E}_{\rm cav} (x, y, t) & = &
\frac {1} {{\cal L}^2} \sum_{\bf k} e^{i (k_x x + k_y y)} {\bf
E}_{{\rm cav}, {\bf k}} (t)
\end{eqnarray}
where ${\bf k} = (k_x,k_y)$ and ${{\cal L}^2}$ is the planar area
of the cavity. Substituting into Eq.~(\ref{E-wave.equ}) we have for
each Fourier component
\begin{equation} \label{E-wave-z.equ}
\frac {n^2_s} {c^2} \frac {\partial^2}{\partial t^2} {\bf {\cal
E}}_{\bf k} + \left( k^2 - \frac {\partial^2}{\partial z^2} \right)
{\bf {\cal E}}_{\bf k} = \frac {\hbar t_c} {\epsilon_0 c^2} \delta
(z) \frac {\partial^2}{\partial t^2} {\bf E}_{{\rm cav}, {\bf k}}
\end{equation}
where $k^2 = k^2_x + k^2_y$. The (positive frequency part of the)
general solution of this equation at any point outside of the
cavity, $z \neq 0$, is given by:
\begin{equation} \label{frequency-comp.equ}
{\bf {\cal E}}_{\bf k} (z,t) = \int^{\infty}_0 \frac {d \omega} {2
\pi} \left[ e^{ i (k_z (k,\omega) z - \omega t)} {\tilde {\bf {\cal
E}}}_{{\bf k},+} (\omega) - e^{i (- k_z (k,\omega) z - \omega t)}
{\tilde {\bf {\cal E}}}_{{\bf k},-} (\omega) \right]
\end{equation}
where $k_z(k, \omega) = + \sqrt {\frac {\omega^2 n^2_s} {c^2} -
k^2}$ and ${\tilde {\bf {\cal E}}}_{{\bf k},+} (\omega)$, ${\tilde
{\bf {\cal E}}}_{{\bf k},-} (\omega)$ are undetermined coefficients.
The negative sign in front of the second term is just a convention.

The unique solution under specified initial/boundary conditions are
obtained by relating the solutions on the two sides of the cavity.
We designate the first integral in Eq.~(\ref{frequency-comp.equ}) as
the right-going field ${\bf {\cal E}}_{\bf k, +} (z,t)$ and the
second integral as the left-going field ${\bf {\cal E}}_{\bf k, -}
(z,t)$. We also put in labels to indicate the domain of the fields:
${\bf {\cal E}}^{R}_{\bf k, +} (z,t)$, ${\bf {\cal E}}^{R}_{\bf k,
-} (z,t)$, ${\tilde {\bf {\cal E}}}^{R}_{{\bf k},+} (\omega)$,
${\tilde {\bf {\cal E}}}^{R}_{{\bf k},-} (\omega)$ are the fields
and coefficients on the right side of the cavity, i.e. for $z > 0$,
and ${\bf {\cal E}}^{L}_{\bf k, +} (z,t)$, ${\bf {\cal E}}^{L}_{\bf
k, -} (z,t)$, ${\tilde {\bf {\cal E}}}^{L}_{{\bf k},+} (\omega)$,
${\tilde {\bf {\cal E}}}^{L}_{{\bf k},-} (\omega)$ are the fields
and coefficients on the left side of the cavity, i.e. for $z < 0$.
We also decompose the oscillator field as
\begin{equation} \label{frequency-osc.equ}
{\bf E}_{{\rm cav}, {\bf k}} (t) = \int^{\infty}_0 \frac {d
\omega} {2 \pi} e^{- i \omega t} {\tilde {\bf E}}_{{\rm cav},
{\bf k}} (\omega)
\end{equation}

The solution fields on the two sides of the cavity are related by
noting that Eq.~(\ref{E-wave-z.equ}) implies continuity of ${\bf
{\cal E}}_{\bf k} (z,t)$ at $z=0$:
\begin{equation} \label{E-continuity.equ}
{\bf {\cal E}}^{R}_{\bf k, +} (0,t) - {\bf {\cal E}}^{R}_{\bf k, -}
(0,t) = {\bf {\cal E}}^{L}_{\bf k, +} (0,t) - {\bf {\cal
E}}^{L}_{\bf k, -} (0,t)
\end{equation}
and a finite jump of $\frac {\partial} {\partial z} {\bf {\cal
E}}_{\bf k} (z,t)$ across $z=0$:
\begin{eqnarray}
&& \frac {\partial} {\partial z} \left[ {\bf {\cal E}}^{R}_{\bf k,
+} (z,t) - {\bf {\cal E}}^{R}_{\bf k, -} (z,t) \right]_{z=0_+} -
\frac {\partial} {\partial z} \left[ {\bf {\cal E}}^{L}_{\bf k, +}
(z,t) - {\bf {\cal E}}^{L}_{\bf k, -} (z,t) \right]_{z=0_-}
\nonumber \\ && \ \ = \ - \frac {\hbar t_c} {\epsilon_0 c^2} \frac
{\partial^2}{\partial t^2} {\bf E}_{{\rm cav}, {\bf k}} (t)
\end{eqnarray}
Using the expansions Eqs.~(\ref{frequency-comp.equ}) and
(\ref{frequency-osc.equ}), we get for each frequency component,
\begin{equation}\label{dE-dz-jump-freq.equ}
{\tilde {\bf {\cal E}}}^{R}_{{\bf k},+} (\omega) + {\tilde {\bf
{\cal E}}}^{R}_{{\bf k},-} (\omega) - {\tilde {\bf {\cal
E}}}^{L}_{{\bf k},+} (\omega) - {\tilde {\bf {\cal E}}}^{L}_{{\bf
k},-} (\omega) = -i \frac {\hbar t_c \omega^2} {\epsilon_0 c^2 k_z}
{\tilde {\bf E}}_{{\rm cav}, {\bf k}} (\omega)
\end{equation}
We again invoke the approximation $k^2 \approx 0 \Rightarrow k_z
\approx \omega n_s / c$ on the right hand side of
Eq.~(\ref{dE-dz-jump-freq.equ}), which becomes $\frac {\hbar t_c}
{\epsilon_0 c n_s} \left[ -i \omega {\tilde {\bf E}}_{{\rm cav},
{\bf k}} (\omega) \right]$.  Taking the inverse Fourier transform
back to the time domain, we get
\begin{equation} \label{dE-dz-jump.equ}
{\bf {\cal E}}^{R}_{\bf k, +} (0,t) + {\bf {\cal E}}^{R}_{\bf k, -}
(0,t) - {\bf {\cal E}}^{L}_{\bf k, +} (0,t) - {\bf {\cal
E}}^{L}_{\bf k, -} (0,t) = \frac {\hbar t_c} {\epsilon_0 c n_s}
\frac {\partial} {\partial t} {\bf E}_{{\rm cav}, {\bf k}} (t)
\end{equation}
Eqs.~(\ref{E-continuity.equ}) and (\ref{dE-dz-jump.equ}) provide
two conditions which allow to express the fields on one side of the
cavity in terms of those on the other side.  The result is
\begin{align}\label{transfer-matrix.equ}
\left(
\begin{array}{c}
{\bf {\cal E}}^{R}_{\bf k, +} (0,t) \\ {\bf {\cal E}}^{R}_{\bf k, -}
(0,t)
\end{array}
\right) = \left(
\begin{array}{c}
{\bf {\cal E}}^{L}_{\bf k, +} (0,t) \\ {\bf {\cal E}}^{L}_{\bf k, -}
(0,t)
\end{array}
\right) + \frac {\hbar t_c} {2 \epsilon_0 c n_s} \frac
{\partial} {\partial t} {\bf E}_{{\rm cav}, {\bf k}} (t) \left(
\begin{array}{c}
1 \\ 1
\end{array}
\right)
\end{align}

In our setting, all the fields have the same circular polarization,
say $+$. The ($+$) polarization vectors for all the slightly
obliquely directed beams are of course not exactly parallel to each
other, but consistent with the small-$k$ approximations we have made
so far, we ignore these slight misalignments and use a common unit
vector ${\hat {\bf e}}_+ = ({\hat x} + i {\hat y})/ \sqrt {2}$.  Then,
using the notations in Ref. \onlinecite{schumacher-etal.09}, we write
${\bf {\cal E}}^{L}_{\bf k, +} (0,t) = {\cal L}^2 E_{{\bf k}, {\rm
inc}} (t) {\hat {\bf e}}_+$, ${\bf {\cal E}}^{L}_{\bf k, -} (0,t) = {\cal
L}^2 E_{{\bf k}, {\rm refl}}(t) {\hat {\bf e}}_+$, ${\bf {\cal
E}}^{R}_{\bf k, +} (0,t) = {\cal L}^2 E_{{\bf k}, {\rm trans}}(t)
{\hat {\bf e}}_+$, ${\bf {\cal E}}^{R}_{\bf k, -} (0,t) = 0$, and ${\bf
E}_{{\rm cav}, {\bf k}} (t) = {\cal L}^2 E_{\bf k} (t) {\hat {\bf e}}_+$.
(We reiterate that $E_{\bf k} (t)$ denotes the cavity oscillator
field and is not to be confused with ${\cal E}_{\bf k} (t)$, which
denotes an macroscopic field mode.) From
Eq.~(\ref{transfer-matrix.equ}), we obtain
\begin{eqnarray}
E_{{\bf k}, {\rm trans}}(t) & = & E_{{\bf k}, {\rm inc}} (t) -
E_{{\bf k}, {\rm refl}}(t) \label{transfer-1.equ} \\ E_{{\bf k},
{\rm refl}}(t) & = & - \frac {\hbar t_c} {2 \epsilon_0 c n_s} \frac
{\partial} {\partial t} E_{\bf k} (t) \label{transfer-2.equ}
\end{eqnarray}
It is also clear that $E_{{\bf k}, {\rm trans}}(t)$ is the radiation
field at the position of the cavity and thus is equal to the driving
radiation field in the equation for $E_{\bf k} (t)$,
Eq.~(\ref{Ek.equ}): $E^{\rm eff}_{{\bf k}, {\rm inc}} = E_{{\bf k},
{\rm trans}}(t)$.

In the remainder of this appendix, we show that the postulated
equation of motion for the cavity field $E_{\bf k} (t)$, Eq.
(\ref{Ek.equ}), leads, in conjunction with the equations for the
macroscopic field and the QW exciton field, naturally to the
interpretation of $| E_{\bf k} (t) |^2$ as the number of photons in
mode $\bf k$ per unit area. Our argument is based on considering the
energy balance in our theory. From the Maxwell equations and the
constitutive equations Eqs. (\ref{D-P.equ}) and (\ref{H-B.equ}), we
obtain in the usual way the continuity equation for the macroscopic
field's energy flux:
\begin{equation}\label{Poynting.equ}
\frac {\partial} {\partial t} u_{\rm em} + \nabla \cdot {\bf S} = +
2 \hbar t_c \delta ( z ) {\rm Re} \left[ {\bf {\cal E}}^\ast \cdot
\frac {\partial} {\partial t} {\bf E}_{\rm cav}  \right]
\end{equation}
where
\begin{eqnarray}
u_{\rm em} & = & n^2_{\rm b} \epsilon_0 {\bf {\cal E}}^\ast \cdot {\bf
{\cal E}} + \frac {1} {\mu_0} {\bf {\cal B}}^\ast \cdot {\bf {\cal B}} \\
{\bf S} & = & 2 {\rm Re} \left[ {\bf {\cal E}}^\ast \times \frac {1}
{\mu_0} {\bf {\cal B}} \right]
\end{eqnarray}
$u_{\rm em}$ is the averaged light
field energy density, including the energy of interaction with the
background dielectric, and $\bf S$, the Poynting vector, is the
averaged energy flux density. Integrating over an
infinitesimal interval across the cavity at $z=0$ gives
\begin{equation}\label{cavity-flux.equ}
\hat{z} \cdot \left( 2 {\rm Re} \left[ {\bf {\cal E}}^\ast \times
\frac {1} {\mu_0} {\bf {\cal B}} \right]_{z=0_+} - 2 {\rm Re} \left[
{\bf {\cal E}}^\ast \times \frac {1} {\mu_0} {\bf {\cal B}}
\right]_{z=0_-} \right) = \hbar t_c 2 {\rm Re} \left[ {\bf {\cal
E}}^\ast (z=0) \cdot \frac {\partial} {\partial t} {\bf E}_{\rm cav}
\right]
\end{equation}
Since the left hand side of Eq. (\ref{cavity-flux.equ}) is the net
field energy flux density coming out of the cavity, the right hand
side should be interpreted as the negative of the rate of energy
change per unit area inside the cavity. If we integrate the right
hand side over the cavity's plane and specialize to our setting, we
get
\begin{eqnarray}
\hbar t_c \int dx dy 2 {\rm Re} \left[ {\bf {\cal E}}^\ast (z=0)
\cdot \frac {\partial} {\partial t} {\bf E}_{\rm cav} \right] & = &
\frac {\hbar t_c} {{\cal L}^2} \sum_{\bf k} 2 {\rm Re} \left[ {\bf
{\cal E}}^\ast_{\bf k} (0,t) \cdot \frac {\partial} {\partial t}
{\bf E}_{{\rm cav},{\bf k}} \right] \nonumber \\ & = & \hbar t_c
{\cal L}^2 \sum_{\bf k} 2 {\rm Re} \left[ E^\ast_{{\bf k},{\rm
trans}}\frac {\partial} {\partial t} E_{\bf k} \right]
\label{cavity-energy-1.equ}
\end{eqnarray}
We now use the equation of motion of $E_{\bf k} (t)$, Eq.
(\ref{Ek.equ}), to eliminate $E^\ast_{{\bf k},{\rm trans}}$, and we
get for the right hand side of Eq. (\ref{cavity-energy-1.equ}):
\begin{equation}\label{cavity-energy-2.equ}
- {\cal L}^2 \sum_{\bf k} \left[ \hbar \omega^c_{\bf k} \frac
{\partial} {\partial t} | E_{\bf k} |^2 - 2 {\rm Re} \left(
\Omega_{\bf k} p^\ast_{\bf k} \frac {\partial} {\partial t}
E_{\bf k}  \right) \right]
\end{equation}
$\Omega_{\bf k}$ being real. From the first term, one can see that
$| E_{\bf k} |^2$ is to be interpreted as the number of photons per
unit area in the cavity mode ${\bf k}$. The second term represents
energy exchange with the exciton field. We can cast it into a more
physically meaningful form by using $p_{\bf k}$'s equation of
motion. For simplicity we illustrate this point with the linearized
form of Eq. (\ref{pk.equ}). We first write the term in question as
\begin{equation}
- 2 {\rm Re} \left( \Omega_{\bf k} p^\ast_{\bf k} \frac
{\partial} {\partial t} E_{\bf k} \right) = - \frac {\partial}
{\partial t} 2 {\rm Re} \left( \Omega_{\bf k} p^\ast_{\bf k}
E_{\bf k} \right) + 2 {\rm Re} \left( \Omega_{\bf k} E_{\bf k}
\frac {\partial} {\partial t} p^\ast_{\bf k}  \right)
\end{equation}
For the second term, we eliminate $E_{\bf k}$ using Eq.
(\ref{pk.equ}) without the nonlinear terms and finally obtain for Eq.
(\ref{cavity-energy-2.equ})
\begin{align}\label{cavity-energy-3.equ}
- {\cal L}^2 \sum_{\bf k} \left\{ \frac {\partial} {\partial t}
\left[
 \left(
\begin{array}{c c}
E^\ast_{\bf k} & p^\ast_{\bf k}
\end{array}
\right) \left(
\begin{array}{c c}
\hbar \omega^c_{\bf k} & - \Omega_{\bf k} \\ - \Omega_{\bf k} &
\varepsilon^x_{\bf k}
\end{array}
\right) \left(
\begin{array}{c}
E_{\bf k} \\ p_{\bf k}
\end{array}
\right) \right] + 2 \gamma_x {\rm Im} \left[ p_{\bf k} \frac
{\partial} {\partial t} p^\ast_{\bf k} \right] \right\}
\end{align}
The first term is the rate of change of energy of the {\em linear}
polariton field inside the cavity, and the second term is the rate
of energy loss by dissipation.

We can also reduce the left hand side of Eq. (\ref{cavity-flux.equ})
to our case where all fields are in the same circular polarization
(spin) state. We state the result here:
\begin{equation}\label{cavity-flux-LHS.equ}
\frac  {c} {n_{\rm b}} {\cal L}^2 \sum_{\bf k} 2 n^2_{\rm b}
\epsilon_0 \left[ | E_{{\bf k}, {\rm trans}}(t) |^2 + | E_{{\bf
k}, {\rm refl}}(t) |^2 - | E_{{\bf k}, {\rm inc}}(t) |^2 \right]
\end{equation}
It is of course also simple to verify the equality between Eq.
(\ref{cavity-flux-LHS.equ}) and Eq. (\ref{cavity-energy-2.equ}) by
directly applying Eqs. (\ref{Eeff.equ})-(\ref{E_eff_trans.equ}).

\newpage

\bibliographystyle{prsty_noetal}

\end{document}